\documentclass[aps,pra,twocolumn]{revtex4}

\usepackage{amsmath}
\usepackage{amssymb}
\usepackage{mathrsfs}
\usepackage{epsfig}
\usepackage{color}
\usepackage[bookmarks]{hyperref}
\usepackage{filecontents}
\usepackage{graphicx}
\usepackage{mathrsfs}
\usepackage{dsfont}
\graphicspath{{.}{./}}
\usepackage{enumerate}
\usepackage{mathtools}
\usepackage{bm}

\usepackage{soul}

\newcommand* {\bra}[1]{\ensuremath{\langle {#1} |}}
\newcommand* {\ket}[1]{\ensuremath{| {#1} \rangle}}

\begin{document}

\title{Estimation of disorders in the rest positions of two membranes in optomechanical systems}

\author{Claudio M. Sanavio}
\affiliation{Department of Physics, University of Bologna, Via Irnerio 46, 40126 Bologna, Italy}

\author{J\'ozsef Zsolt Bern\'ad}
\affiliation{Peter Gr\"unberg Institute (PGI-8), Forschungszentrum J\"ulich, D-52425 J\"ulich, Germany}

\author{Andr\'e Xuereb}
\affiliation{Department of Physics, University of Malta, Msida MSD 2080, Malta}

\date{\today}

\begin{abstract}
The formalism of quantum estimation theory is applied to estimate the disorders in the positions of two membranes positioned in a driven optical cavity. We consider the coupled-cavities and the transmissive-regime
models to obtain effective descriptions of this system for different reflectivity values of the membranes. Our models consist also of high temperatures Brownian motions of the membranes, losses of the cavity 
fields, the input-output formalism, and a balanced homodyne photodetection of the cavity output field. In this two-parameter estimation scenario, we compare the classical and quantum Fisher information matrices 
and evaluate the accuracies of the estimations. We show that models prefer very different estimation strategies and the temperature does not have a detrimental effect on the estimation accuracies but makes it more 
difficult to attain the quantum optimal limit. Our analysis, based on recent experimental parameter values, also reveals that the best estimation strategies with unit efficient detectors are measurements of the quadratures of the output field.

\end{abstract}

\maketitle

\section{Introduction}\label{sec:I}

Parameter estimation is a crucial task at the heart of engineering and physical sciences \cite{Kaipio}. Quantum statistical inference attempts to find appropriate quantum measurements or estimators, from which
the value of one or more parameters of a quantum mechanical system can be estimated \cite{Helstrom, Holevo, Wiseman}. This task may not always guarantee implementable measurements with current technologies, and 
therefore one has to consider a family of quantum measurements used in recent experimental setups. These measurements generate data that is inherently random, it is usually described by a 
probability density function depending on the true values of the parameters to be estimated. Estimators are functions on the data and their performance are usually assessed by their mean-squared error or variance 
when they are unbiased. Being able to place a lower bound on the mean-squared error or variance of any estimator provides us a benchmark against which we can compare the performances of different estimation strategies. 
Although many lower bounds exist for classical systems \cite{KBell}, Cram\'er-Rao lower bound is the one which has a straight extension to quantum systems and is by far the easiest to determine \cite{Helstrom68}. 
In the multi-parameter estimation case with unbiased estimators, which is our intention here, the covariance matrix of the estimates is lower bounded by the inverse of the 
quantum Fisher information matrix (QFIM) in terms of matrix inequalities. Provided that we would like to perform inference in a quantum mechanical system with a constrained set of quantum measurements, 
the process of estimation is divided in our approach into two parts. First, one determines the classical Fisher information matrix (CFIM) from the  probability density function of the measurement data and 
investigates circumstances where the CFIM is in the trace norm as close as possible to the QFIM, which in terms of matrix inequalities is always larger or equal than the CFIM \cite{Petz}. 
Finally, in the classical postprocessing of measurement data, the attainability of the Cram\'er-Rao lower bound is investigated. \cite{vanTrees}. 

In this paper, we follow the above-described methodology for estimating the disorder in the positions of mechanical membranes in an optical cavity. Multiple-membrane cavity optomechanics is getting increasing 
attention from the scientific community in the last decade. In contrast with the standard optomechanical set-up of a linear cavity composed of one fixed mirror and one movable end mirror, 
the membrane-in-the-middle (MIM) 
configuration sees the movable membrane, a dielectric thick surface, in between the two fixed mirrors composing the optical cavity. The interesting features of this set-up have been investigated both theoretically 
\cite{Chow1986,BhattacharyaMeystre2007,BhattacharyaUysMeystre2008} and experimentally \cite{Jayich2008}. The presence of the dielectric material changes the properties of the optical mode, 
its frequency, and therefore the position of nodes and anti-nodes. Following these interesting results, the theoretical investigation had shifted to multiple membrane-in-the-middle (MMIM) configuration \cite{BhattacharyaMeystre2008}, where more membranes are 
located inside the optical cavity. The analysis of these systems showed promising features, like the enhancement of optomechanical coupling strengths based on constructive
interference \cite{XuerebGenesDantan2012,Rabl2011,LiXuerebMalossiVitali2016}. 

Optomechanical systems are well-suited for studying the nature of quantum mechanics of macroscopic objects \cite{Marshall} as well as measuring weak forces with high sensitivity and precision \cite{Kippenberg}. 
They lie at the heart of laser-based interferometric gravitational wave observatories \cite{Abbott}, the theoretical background of which has been known for several decades \cite{Braginsky,Caves}. 
In such systems, the physical quantity of interest is encoded in the displacement of the moving element, which must be estimated with the greatest possible precision. These considerations carry forward to the 
MMIM scenario, and in particular to systems with two moving membranes, of which experimental investigations started only recently \cite{Piergentili2018,Wei2019}. 
However, previously set rest positions of the membranes can be displaced due to imperfections, and hence the precision of the whole experimental setup is affected. Here, we provide a systematic estimation of 
disorders in the positions 
of the membranes based on statistical inference. We follow and extend our previous frequentist statistical inference approach \cite{SanavioBernadXuereb2020}, where the measurement data on a cavity optomechanical
system is obtained by determining the output field of the cavity with the help of the input-output relations \cite{Gardiner1985} and measuring the escaping field by balanced homodyne photodetection.

This paper is organized as follows. In Sec.~\ref{sec:II} we introduce two models describing the system when the reflectivity of the membranes is either high or low. Then, we employ the Heisenberg-Langevin equations
to obtain the steady-state and its fluctuations for the output field to be measured. In Sec.~\ref{sec:III}  we discuss our multi-parameter estimation strategy in the context of 
balanced homodyne photodetection. Then, we apply our strategy to infer the disorder in the positions of the membranes. In Sec.~\ref{sec:IV} we show the results and in Sec.~\ref{sec:V} we draw our conclusions. 
Detailed formulas supporting the main text are collected in the four appendices.

\section{Model}\label{sec:II}

\begin{figure}
 \includegraphics[width=.39\textwidth]{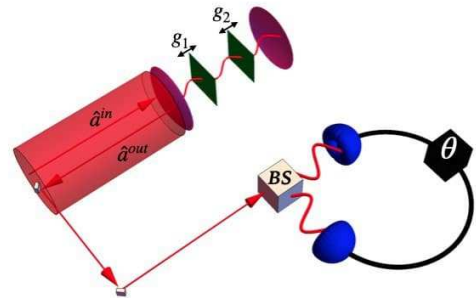}
\caption{Schematic representation of the system. Two movable dielectric membranes are placed inside a Fabry–P\'erot cavity. Further details about the scheme are in the text.}
   \label{fig:model}
\end{figure}

We consider an optical cavity of length $L$ formed by perfectly reflecting end mirrors and two identical vibrating dielectric membranes, which are placed inside the cavity (see. Fig. \ref{fig:model}). 
Each of these membranes has reflectivity $r$, mass $m$, and mechanical frequency $\omega_m$. Furthermore, they are bounded by a harmonic potential $m\omega_m^2\hat{q}^2_i/2$, 
with $\hat{q}_i$ being the position operator.

One is able to find the electromagnetic field inside the cavity by solving the Helmholtz equation and setting the proper boundary conditions~\cite{Brooker2003}. However, the electric 
susceptibility inside the full cavity has to be modeled in order to incorporate both membranes~\cite{BhattacharyaMeystre2008}. 
In order to make the canonical quantization of such a  system possible, Ref.~\cite{CheungLaw2011} has assumed a nonbirefringent membrane, i.e, refractive
index $r$ does not depend on the polarization and propagation direction of the field, and also a nondispersive one, i.e, electric susceptibility of the membrane does not depend on the field's frequency. 
Now, based on the single membrane approach of Ref.~\cite{CheungLaw2011} we consider an identical second membrane. We assume that the two membranes have independent suspensions and therefore the second membrane 
is modeled as an additive contribution to the Hamiltonian of Ref. \cite{CheungLaw2011}. If the harmonic potentials bound both dielectric membranes about their rest positions such that the average position operator$\langle\hat{q}_i\rangle, (i = 1, 2)$ is small 
compared to the wavelengths of the field, then the linear approximation of field-membrane couplings is valid and the Hamiltonian reads
\begin{eqnarray}\label{eq:Hamiltonian1}
\hat{H}&=&\sum_{j}\hbar\omega_{j}\hat{a}_{j}^\dagger\hat{a}_{j}+\sum_{i=1}^2\hat{q}_i\sum_{j,k} g_{ijk}\bigg{(}\hat{a}_{j}^\dagger\hat{a}_{k}^\dagger+\hat{a}_{j}^\dagger\hat{a}_{k}+\text{h.c.}\bigg{)}\nonumber\\
&&+\sum_{i=1}^2\bigg{[}\frac{\hat{p}_i^2}{2m}+\frac{m\omega_\text{m}^2 \hat{q}^2_i}{2}\bigg{]}.
\end{eqnarray}
where $\hat{a}_j$ $(\hat{a}^\dagger_j)$ is the annihilation (creation) operator of the $j$th field mode with frequency $\omega_j$, which is obtained in the case when both membranes are in rest. 
Similarly, the coupling constants $g_{ijk}$ are \cite{CheungLaw2011}:
\begin{equation}
 g_{ijk}=\frac{\partial \omega_j(q)}{\partial q} \big |_{q=q^{(0)}_i} \delta_{j,k}+J_{ijk}, \label{eq:defgJ}
\end{equation}
where $\delta_{j,k}$ is the Kronecker delta and $q_i^{(0)}$s $(i=1,2)$ are the rest positions of the membranes in the absence of the electromagnetic field.  $J_{ijk}$ is the coupling strength of photon 
emission and absorption processes, which occur between field modes $j$ and $k$ and are mediated by the $i$th membrane. We have also the following
property: $J_{ijj}=0$ for all $i$ and $j$. 

In this paper, we are interested in two different setups, in which differences are marked by the reflectivity of the membranes.
When the reflectivity is high, one can use the so-called coupled-cavities (CC) model, for which three different modes are localized in the spacings between membranes and end mirrors. 
On the other hand, when the reflectivity is low 
we consider the electromagnetic field mode to be delocalized in the cavity. We dub this model the Transmissive regime (TR). 
Both CC and TR models have been largely investigated in literature \cite{BhattacharyaMeystre2008,Jayich2008,XuerebGenesDantan2013} and represent the two most pursued effective models for multiple optomechanical systems. Our goal is to show the differences and analogies during an estimation process.

\subsection{Dissipative dynamics}\label{sec:IIa} 

Membranes interact with the
surrounding gas atoms and are also coupled to the environment through the suspensions. Their dynamics are slow compared to the correlation times of the environments. This is the characteristic case of 
quantum Brownian motion and without loss of generality, we consider this as the only dissipative mechanism of the membranes, though, loss of mechanical excitations is a rich phenomenon. 
\cite{AspelmeyerKippenbergMarquardt2014}. 
Quantum Brownian motion in a harmonic potential $m\omega_\text{m}^2 \hat{q}^2/2$ is described by the following Heisenberg equations of motion
\begin{eqnarray}
 \dot{\hat{q}}&=&\frac{\hat{p}}{m}, \label{eq:QLEBrown} \\
\dot{\hat{p}}&=&-m\omega_\text{m}^2 \hat{q}-\gamma\hat{p}+\hat{\xi} .\nonumber
\end{eqnarray}
where $\gamma$ is the strength of the friction force. The operator $\hat{\xi}$ represents the quantum Brownian noise and we consider that the environment was initially in a thermal equilibrium state 
with temperature $T$. In the high-temperature limit, which is valid at room temperatures, the two-time correlation function of $\hat{\xi}(t)$ reads \cite{BreuerPetruccioni2002}:
\begin{eqnarray}
\langle\hat{\xi}(t)\hat{\xi}(t')\rangle=2 m \gamma k_B T \delta(t-t'). \nonumber
\end{eqnarray}

Any mode of the field inside the optical cavity is subject to photon leakage through mirrors and membranes, which couple the inside field with the continuum of the outside field modes.
The dynamic of any optical cavity mode is well described by the Heisenberg-Langevin equation~\cite{Gardiner1991}. Based on the input-output formalism this equation is given by the time evolution 
of the single mode field operator $\hat{a}$ subject to decay $\kappa$ and affected by noise, which appears explicitly as the input field $\hat{a}^{\text{in}}$. This equation reads
\begin{equation}\label{eq:QLEopt}
\frac{d\hat{a}}{dt}=-\frac{\kappa}{2}\hat{a}+\sqrt{\kappa}\hat{a}^{\text{in}},
\end{equation}
where we have omitted, for now, the full Hamiltonian evolution of the system.
The input operator $\hat{a}^{\text{in}}$ associated with the
vacuum fluctuations of the continuum of modes outside the cavity is delta correlated in the vacuum state $\bra{0}[\hat{a}^{\text{in}}(t),\hat{a}^{\dagger\text{in}}(t')]\ket{0}=\delta(t-t')$.
This is because the field modes have optical frequencies and thus the average number of thermal photons for these frequencies at room temperature is approximately zero. 
Furthermore, we can use the same input-output theory to describe the losses induced by the manufacturing errors of the membranes. As a result, in the CC model, each subcavity can experience 
a different decay rate $\kappa_j$ ($j\in\{1,2,3\}$). In the case of the TR model, we consider only one decay rate for the single mode field. 

\subsection{Coupled-cavities (CC)}\label{sec:IIb}

When two membranes are placed inside an optical cavity, then there are three spacing or inner cavities between the membranes and the mirrors. We denote the length of each inner cavity by $L/3$ 
and thus the difference between the rest positions of the membranes $q^{(0)}_2-q^{(0)}_1=L/3$. If the reflectivity of the membranes is one, i.e., $r=1$, and they are resting, the optical cavity consists simply of three uncoupled inner cavities with eigenfrequencies
\begin{equation}\label{eq:CCFrequencies}
\omega_{n}=\frac{3 n\pi c}{L},
\end{equation}
where $n$ is a positive integer and $c$ is the speed of light. Provided that the reflectivity is slightly smaller than one then the three inner cavities become coupled. 
Furthermore, we consider that in each inner cavity only one field mode is dominant with frequency $\omega_c$. This condition can be achieved by driving the system with a laser such that 
only these selected modes are enhanced \cite{BhattacharyaMeystre2008}. Therefore, we consider a laser with frequency $\omega_\text{L}$ and intensity $\varepsilon$ driving the first inner cavity ($j=1$) 
having a mirror for the left and a membrane for the right boundary. Now, if in \eqref{eq:Hamiltonian1} we neglect scattering processes between dominant and non-dominant modes and also the two-photon 
processes, we obtain 
\begin{eqnarray}\label{eq:CCHamiltonian}
H_{\text{CC}}&=& \sum\limits_{j=2}^{3}\hbar\Delta_0 \hat{a}_j^\dagger\hat{a}_j+i\hbar \varepsilon(\hat{a}_1^{\dagger}-\hat{a}_1)\nonumber\\
&+&\sum\limits_{i=1}^{2}\bigg{[}\hbar g\hat{q}_i(\hat{a}^{\dagger}_{i}\hat{a}_i-\hat{a}^{\dagger}_{i+1}\hat{a}_{i+1}) \\
&+&\frac{\hat{p}^2_i}{2m}+\frac{m\omega_\text{m}^2 \hat{q}^2_i}{2}+\hbar J\hat{a}_{i+1}^\dagger\hat{a}_i+\hbar J\hat{a}_{i}^\dagger\hat{a}_{i+1}\bigg{]}, \nonumber
\end{eqnarray}
where $\Delta_0=\omega_c-\omega_\text{L}$, $g=g_{111}=g_{122}=g_{222}=g_{233}$ and $J=g_{112}=g_{121}=g_{223}=g_{232}$, see Eq. \eqref{eq:defgJ}. Note that Hamiltonian~\eqref{eq:CCHamiltonian} is 
already expressed in a rotating-frame of all three modes of the field with frequency $\omega_{\text{L}}$ \cite{AspelmeyerKippenbergMarquardt2014}. 

The $i$th and $(i+1)$th modes are located at the left and right of the $i$th ($i \in \{1,2\}$) membrane respectively and therefore they exert an 
opposite light pressure on this membrane, which is reflected on the different signs of the coupling between the mechanical motion of the membrane and the two adjacent single mode fields. 
The electromagnetic field either passes through or pushes the membranes, where these effects are characterized by the hopping rate $J$ and uniform optomechanical coupling $g$. Both processes influence the motion
of the membranes and thus the Hamiltonian can describe rich physics, though, due to the number of assumptions involved is still a ``minimal''-model. In fact, the optical hopping between the inner 
cavities accounts for the non-perfect reflectivity of the membranes, and for example, a classical understanding of the hopping rate with the method of the transfer matrix yields the relation 
$J=\omega_c \sqrt{2(1-r)}$~\cite{Jayich2008}.  

A standard procedure consists of linearizing the dynamics by expanding the Hamiltonian around the steady-state~\cite{AspelmeyerKippenbergMarquardt2014}, which is reached due to decoherence and 
excitation losses in the system, see Sec. \ref{sec:IIa} for further details. This procedure is defined through the transformations $\hat{a}_j\to\alpha_j+\hat{a}_j$ and $\hat{q}_i\to q_{i}+\hat{q}_i$, 
where $\alpha_j$ and $q_{i}$ represent the steady state solutions for the $j$th field mode and the $i$th membrane respectively. The transformation is applied to the dynamics of the system, 
(see Appendix \ref{app:steadystateamplitude}), and doesn't affect the momentum operators $\hat{p}_i$ \cite{SanavioBernadXuereb2020}. The new operators describe the fluctuations around the steady-state and 
second-order terms in the transformed Hamiltonian are neglected. Finally, we can find the Hamiltonian that rules the dynamics of the fluctuation operators, which reads
\begin{eqnarray}\label{eq:linearCCHamiltonian}
H_{\text{CC,quad}}&=&\sum_{j=1}^{3}\hbar\Delta_j\hat{a}_j^\dagger\hat{a}_j+\sum_{i=1}^{2}\bigg{(}\hbar J\hat{a}_{i+1}^\dagger\hat{a}_i+\hbar J\hat{a}_{i}^\dagger\hat{a}_{i+1}\nonumber\\
&&+\hbar g\hat{q}_i(\alpha_i\hat{a}^{\dagger}_{i}-\alpha_{i+1}\hat{a}^{\dagger}_{i+1}+\bar{\alpha}_i\hat{a}_{i}-\bar{\alpha}_{i+1}\hat{a}_{i+1})\nonumber\\
&&+\frac{\hat{p}^2_i}{2m}+\frac{m\omega_\text{m}^2 \hat{q}^2_i}{2}\bigg{)},
\end{eqnarray}
where the detunings $\Delta_j$ and the steady-state amplitudes $\alpha_j$ for the $j$th ($j \in \{1,2,3\}$) inner cavity can be obtained as a function of parameters of the Hamiltonian and  
loss mechanisms. We denoted $\bar{\alpha}$ as the complex conjugate of $\alpha$. We use this quadratic Hamiltonian to calculate the dynamics of the fluctuations and estimate the disorder in the positions 
of the membranes.

\subsection{Transmissive regime (TR)}\label{sec:IIc}

We now consider a different situation, where only one mode is present in the whole cavity and interacts with each of the membranes. This can be realized in what is called the transmissive regime 
\cite{XuerebGenesDantan2012} of the membrane stack. In general, for any value $r$ of a $N_m$ membranes system, one can find $N_m+1$ selected lengths $L/(N_m+1)$ of the inner cavities, 
such that the global reflectivity of the whole membrane set drops to zero. Thus, the field sees the membrane stack as a single membrane with low reflectivity, regardless of the original value of $r$.
An analytical expression for the different optomechanical coupling strengths can also be obtained by using the transfer matrix method \cite{Piergentili2018,XuerebGenesDantan2013}. This is our starting point, 
where we consider a single mode field  with frequency $\omega_c$ in the whole optical cavity. A laser with frequency $\omega_\text{L}$ and intensity 
$\varepsilon$ is also driving this mode. Now, we obtain another subcase of \eqref{eq:Hamiltonian1}, which reads
\begin{eqnarray}\label{eq:HamiltonianOneMode}
H_{\text{TR}}&=&\hbar\left(\Delta_0+\sum^2_{i=1} g_i \hat{q}_i\right)\hat{a}^\dagger\hat{a}\nonumber\\
&&+\sum^2_{i=1}\bigg{(}\frac{\hat{p}_i^2}{2m}+\frac{m\omega_\text{m}^2\hat{q}^2_i}{2}\bigg{)}\nonumber\\
&&+i \hbar \varepsilon(\hat{a}^{\dagger}-\hat{a}),
\end{eqnarray}
where $\Delta_0=\omega_c-\omega_\text{L}$, $g_1=g_{111}$, and $g_2=g_{211}$, see Eq. \eqref{eq:defgJ}. We immediately went to the rotating-frame
of the field mode with frequency $\omega_\text{L}$ and assumed  the disorder-free optomechanical coupling strengths of both membranes are equal.

The Hamiltonian which describes the dynamics in terms of the fluctuation is immediate. Based on the arguments of Sec. \ref{sec:IIb} this Hamiltonian in the transmissive regime reads
\begin{eqnarray}\label{eq:HamiltonianOneModeLinear}
H_{\text{TR,quad}}&=&\hbar\left[\Delta\hat{a}^\dagger\hat{a}+\sum^2_{i=1} g_i \hat{q}_i(\alpha \hat{a}^\dagger+\bar{\alpha}\hat{a})\right]\nonumber\\
&&+\sum^2_{i=1}\bigg{(}\frac{\hat{p}_i^2}{2m}+\frac{m\omega_\text{m}^2 \hat{q}^2_i}{2}\bigg{)},
\end{eqnarray}
where the equations yielding the detuning $\Delta$ and the steady-state amplitude $\alpha$ are found in Appendix \ref{app:BCCmodel}.

\subsection{Heisenberg-Langevin equations}\label{sec:IId}

In the following, we present a general formalism that applies to both quadratic Hamiltonians in \eqref{eq:linearCCHamiltonian} and \eqref{eq:HamiltonianOneModeLinear}.
We collect the operators of both dynamics into vectors of operators
\begin{eqnarray}
 c^{(\text{CC})}&=&(\hat{X}_1,\hat{Y}_1, \hat{X}_2, \hat{Y}_2,\hat{X}_3,\hat{Y}_3,\hat{p}_1,\hat{p}_2,\hat{q}_1,\hat{q}_2)^T, \label{CCc}\\
 c^{(\text{TR})}&=&(\hat{X}, \hat{Y}, \hat{p}_1,\hat{p}_2,\hat{q}_1,\hat{q}_2)^T, \label{TRc}
\end{eqnarray}
where the superscript $T$ denotes the transposition and we have introduced the quadratures $\hat X=(\hat a^\dagger + \hat a)/\sqrt{2}$ and $\hat Y=i(\hat a^\dagger - \hat a)/\sqrt{2}$.
We write the corresponding Heisenberg-Langevin equation as
\begin{eqnarray}\label{eq:vectorialQLE}
\frac{d}{dt} c^{(m)}&=&A^{(m)}c^{(m)}+\eta^{(m)}, \quad m\in \{\text{CC},\text{TR}\}, \nonumber \\
\end{eqnarray}
where $\eta^{(m)}$ is the vector of all noise operators:
\begin{eqnarray}
 \eta^{(\text{CC})}&=&(\sqrt{\kappa_1} \hat{X}^{\text{in}}_1,\dots,\sqrt{\kappa_3}\hat{Y}^{\text{in}}_3,\hat{\xi},\hat{\xi},0,0)^T, \nonumber \\
 \eta^{(\text{TR})}&=&(\sqrt{\kappa} \hat{X}^{\text{in}},\sqrt{\kappa}\hat{Y}^{\text{in}},\hat{\xi},\hat{\xi},0,0)^T. \nonumber
\end{eqnarray}
The dynamical matrices $A^{(CC)}$ and $A^{(TR)}$ contain terms obtained from the quadratic Hamiltonians and their explicit forms can be 
found in Appendices \ref{app:steadystateamplitude} and \ref{app:BCCmodel}. Finally, the formal solution of \eqref{eq:vectorialQLE} reads
\begin{eqnarray}\label{eq:QLEsolution}
c^{(m)}(t)&=&\exp\left[A^{(m)}t\right]c^{(m)}(0) \\
&+&\int_0^tdt'\exp\left[A^{(m)}(t-t')\right]\eta^{(m)}(t'). \nonumber
\end{eqnarray}

The quadratic Hamiltonians in \eqref{eq:linearCCHamiltonian} and \eqref{eq:HamiltonianOneModeLinear} together with the 
loss mechanisms ensure that the state of the fluctuations is Gaussian~
\cite{WeedbrookPirandola2011}. As the fluctuations around the steady-state have zero means, it is immediate that this Gaussian state
is fully described by the symmetric auto-correlation matrix 
\begin{equation}
\sigma(t,t')=\frac{1}{2}\langle c(t) c(t')^T+ c(t')c(t)^T\rangle.
\end{equation}

We use the solutions of the quantum Langevin equations \eqref{eq:QLEsolution} to find $\sigma(t,t)$ in the stationary limit $t \to \infty$. Provided that the 
both systems are stable, where conditions are derived by using the Routh-Hurwitz criterion \cite{Routh}, $\sigma=\lim_{t \to \infty} \sigma(t,t)$ fulfills the following
Lyapunov equation~\cite{SanavioBernadXuereb2020}.
\begin{eqnarray}\label{eq:Lyapunov}
A\sigma+\sigma A^T=-D,
\end{eqnarray}
where
\begin{equation}\label{eq:DMatrix}
D=\int_0^{\infty} d\tau[M(\tau)\exp(A^T\tau)+\exp(A\tau)M(\tau)],
\end{equation}
and
\begin{equation}
M(t-t')=\frac{1}{2}\langle \eta(t)\eta(t')^T+\eta(t')\eta(t)^T\rangle
\end{equation} 
is noise correlation matrix. In particular, the matrix entries are:
\begin{eqnarray}\label{eq:MMatrix}
\left[M\right]_{\hat{X}_i\hat{X}_j}(t-t')&=&\left[M \right]_{\hat{Y}_i\hat{Y}_j}(t-t')= \frac{\kappa_i}{2}\delta_{i,j}\delta(t-t')\nonumber\\
\left[M\right]_{\hat{p}_i\hat{p}_j}(t-t')&=&2m\gamma k_{\text{B}}T\delta_{i,j}\delta(t-t'). \nonumber
\end{eqnarray} 

Any experiment seeking to infer one or more parameters of this system has to perform measurements on the cavity output field. 
With the help of the input-output relations and considering that the output field possesses the same correlation functions as the optical input field, we have
\begin{equation}
\hat{a}^{\text{out}}=\sqrt{\kappa}\hat{a}-\hat{a}^{\text{in}},
\end{equation}
from which we can find the output correlation matrix $\sigma^{\text{out}}$. 

As the measurement is performed in a finite time interval $\tau$, only some frequencies are accessible to a detector. Hence, we can define the filter function $g_l(t)$~\cite{GenesMariTombesiVitali2008}, 
which accounts for a finite period of detection and is
\begin{equation}
g_l(t)=\frac{\theta(t)-\theta(t-\tau)}{\sqrt{\tau}}e^{-i\Omega_l t},
\end{equation}
with $\Omega_i-\Omega_j=\frac{2\pi}{\tau}n$ and $n\in\mathbb{N}$. The latter condition allows us to define $N$ independent output modes
\begin{equation}
a_l^{\text{out}}(t)=\int_{-\infty}^t ds g_l(t-s)\hat{a}^{\text{out}}(s), \quad l=1,\dots,N,
\end{equation}
which are centered at the frequency $\Omega_l$ and with bandwidth $1/\tau$. Following our previous results in ~\cite{SanavioBernadXuereb2020,comment},
one can obtain the entries of the $2\times2$ correlation matrix 
$\sigma_l^{\text{out}}$ as
\begin{eqnarray}\label{correlationmatrixoutput1}
\sigma^{\text{out}}_{l,XX}&=&\frac{1}{2} \kappa  \tau  \text{sinc}^2\left(\frac{\Omega_l \tau}{2}\right) \left[\left(\sigma_{XX}-\sigma_{YY}\right) 
\cos (\Omega_l \tau) \right.\nonumber \\
&+&\left. \sigma_{XX}+2 \sigma_{XY} \sin (\Omega_l \tau)+\sigma_{YY}\right]+\frac{1}{2}  \\ \label{correlationmatrixoutput2}
\sigma^{\text{out}}_{l,XY}&=&\frac{1}{2} \kappa  \tau  \text{sinc}\left(\frac{\Omega_l \tau}{2}\right)^2 \left[ \left(\sigma_{YY}-\sigma_{XX}\right) 
\sin ( \Omega_k \tau ) \right. \nonumber \\
&+& \left. 2 \sigma_{XY} \cos (\Omega_l \tau) \right]\\\label{correlationmatrixoutput3}
\sigma^{\text{out}}_{l,YY}&=&\frac{1}{2} \kappa  \tau  \text{sinc}^2\left(\frac{\Omega_l \tau}{2}\right) \left[ \left( \sigma_{YY}- \sigma_{XX}\right) 
\cos (\Omega_l \tau)  \right. \nonumber \\
&+& \left. \sigma_{XX}-2  \sigma_{XY} \sin (\Omega_l \tau)+ \sigma_{YY}\right]+\frac{1}{2},
\end{eqnarray}
where $\sigma_{AB}=\left \langle \hat{A} \hat{B}\right \rangle $ ($A,B \in \{X,Y\}$) are the entries of matrix $\sigma$ and
$\text{sinc}(x)$ is the unnormalized sinc function $\text{sinc}(x)=\sin(x)/x$. In the TR model, $\sigma_{XX}$, $ \sigma_{XY}$, and  $\sigma_{YY}$ are obtained directly from \eqref{eq:Lyapunov},
because there is only one mode of the field. The situation in the CC model is different, the output field will leak from the last ($j=3$) inner cavity and after solving the corresponding Lyapunov
equation $\sigma_{X_3X_3}$, $ \sigma_{X_3Y_3}$, and  $\sigma_{Y_3Y_3}$ have to substituted into Eqs. \eqref{correlationmatrixoutput1}, \eqref{correlationmatrixoutput2}, and \eqref{correlationmatrixoutput3}
to obtain $\sigma_l^{\text{out}}$.

Thus, the state of output field fluctuations is given by the Gaussian Wigner function
\begin{equation}\label{Wigner}
W(\xi)=\frac{1}{2 \pi \sqrt{\det(\sigma^{\text{out}}_l)}}e^{-\frac{1}{2}R^T \left[\sigma^{\text{out}}_l\right]^{-1} R},
\end{equation}
where $R=(X^{\text{out}}_l,Y^{\text{out}}_l)^T$. 

\subsection{Effects of disorder in the positions of the membranes}\label{sec:IIe}

In the following, we are going to present effective versions of both models by assuming that the shift of the equilibrium position from $q_i^{(0)}$ to $q_i=q_i^{(0)}+ \delta q_i$ ($i\in\{1,2\}$)
affects only two main parameters, the frequencies of the field modes and the optomechanical couplings. Provided that $\delta q_i \ll L$, the resonance frequency $\omega_c$ of all three inner cavities in the 
CC model changes as \cite{BhattacharyaUysMeystre2008}
\begin{equation}
\omega_{j}\approx \omega_c \left[1-3\left(\delta q_j-\delta q_{j-1}\right)/L\right], \nonumber
\end{equation}
where $j \in \{1,2,3\}$ and $\delta q_{0}=\delta q_3=0$, because the end mirrors are assumed to not change their positions. There is only one field mode in the case of the TR model, which changes
according to the following function \cite{Piergentili2018}
\begin{equation}
 \omega_c(q_1,q_2)=\frac{n\pi c}{L}+(-1)^n \frac{c}{L} \arcsin \left[F(q_1, q_2) \right]- \frac{c}{L} \theta (q_1, q_2), \label{eq:piergentiliformula}
\end{equation}
where
\begin{eqnarray}
 F(q_1, q_2)&=&\frac{2 \sqrt{r} \cos \left[k(q_1+ q_2)\right] \sin \left[k (q_2-q_1)\right] }{\sqrt{1+r^2-2r\cos\left[2 k (q_2-q_1)\right]}}, \nonumber \\
 \theta (q_1, q_2)&=& \arcsin \left[ \frac{ \sqrt{r} \sin \left[2k (q_2-q_1)\right] }{\sqrt{1+r^2-2r\cos\left[2 k (q_2-q_1)\right]}}\right], \nonumber 
\end{eqnarray}
$k=n \pi/L$ and $n$ is fixed such that the above formula yields the cavity mode with frequency $\omega_c$ when $\delta q_1=\delta q_2=0$. It is worth noting that the phase related to
the reflection of the membranes is set here to zero \cite{Piergentili2018}.

The optomechanical couplings strength is the derivative of the optical mode frequencies at the position of the $i$th membrane $q_i$, see Eq. \eqref{eq:defgJ}. In the case of the CC model, the 
optomechanical couplings of both membranes are changed to \cite{BhattacharyaUysMeystre2008}
\begin{equation}
 g_i(q_i)=\frac{n\pi c}{L^2}\frac{\sqrt{r} \sin \left(2kq_i\right)}{\sqrt{1-r \cos^2 \left(2kq_i\right) }}, \quad i\in\{1,2\}, \nonumber
\end{equation}
where $g_1(q_1^{(0)})=g_2(q_2^{(0)})=g$ and we have assumed that the disorder in the position of one of the membranes on the optomechanical coupling strength of the other membrane is negligible. Furthermore, we consider 
that the mode functions of each field mode in the three cavities are not changed significantly and thus the membrane induced coupling $J$ also remains unaffected \cite{CheungLaw2011}. Finally, in the TR model using
Eq. \eqref{eq:piergentiliformula} the new optomechanical couplings are
\begin{equation}
 g_i=\frac{\partial \omega_c(q_1,q_2)}{\partial q_i} \big |_{q_i=q^{(0)}_i+\delta q_i}, \quad i\in\{1,2\},\nonumber
\end{equation}
and when $\delta q_1=\delta q_2=0$ then we reobtain the optomechanical couplings $g_1$ and $g_2$. 

Therefore, in both models the steady state solutions will also depend on $\delta q_1$ and $\delta q_2$, which have to also be taken into account in the dynamical matrices $A^{(CC)}$ and $A^{(TR)}$, see 
Appendices \ref{app:steadystateamplitude} and \ref{app:BCCmodel}. 

\section{Estimation}
\label{sec:III}

In this section, we employ an estimation strategy concerning the inference of the disorders $\delta q_1$ and $\delta q_2$. Our starting point is the family of Wigner functions 
$W(\delta q)$ with $\delta q=(\delta q_1,\delta q_2)^T$ in 
Eq. \eqref{Wigner} that describes the possible states of the output field. In general, estimation aims to produce estimates of the unknown disorders from repeated
measurements. These measurements are constrained by current technologies, which from the mathematical point of view means that we have access only to a subset of all possible 
positive-operator valued measures (POVM). A lower bound on the variance of any unbiased estimator is given by the Cram\'er-Rao inequality for both classical and quantum systems.
Best-unbiased estimators are those, who can attain this bound. Finding the best-unbiased estimator, which might not even exist, is not an easy task, nonetheless when we also include the reduced number 
of implementable measurements, i.e., the case of our investigation. Therefore, given a set of measurements with tunable parameters, the best-unbiased estimators will be then those whose covariance matrix in a 
properly chosen norm gets close to the Cram\'er-Rao lower bound.

An outline of our view on the estimation approach is the following:
\begin{itemize}
 \item An output field of the cavity is subject to balanced homodyne photodetection (BHD). Based on our theoretical model these measurements provide us a probability density function (PDF), which 
 is functionally dependent on $\delta q$. 
 \item Then, we investigate the circumstances, where the classical Fisher information is the closest to its upper bound or benchmark value, i.e., the 
 quantum Fisher information. This step will set the values of the experimentally tunable parameters and thus providing the best PDF.
 \item After obtaining the best PDF out of BHD, one has to do classical postprocessing of measurement data. As soon as the PDF is known and the measurement data is available, 
 a standard decision-making process of finding the best classical estimator is carried out.
\end{itemize}

In our two-parameter estimation scenario, the covariance matrix $C(\delta q)$ of the estimates
$\delta q=(\delta q_1,\delta q_2)^T$ fulfills \cite{Petz}
\begin{equation}\label{eq:inequality}
C(\delta q) \geq \mathcal{F}^{-1} \geq \mathcal{H}^{-1},
\end{equation} 
in terms of matrix inequalities, where $\mathcal{F}$ and $\mathcal{H}$ are the classical and quantum Fisher information matrices, respectively. In this sense the 
difference matrix $\mathcal{F}^{-1}-\mathcal{H}^{-1}$ is always non-negative definite. 

The quantum Fisher information matrix (QFIM) depends only on the family of states $\rho(\delta q)$ and its components are
\begin{equation}
\mathcal{H}_{ij}=\frac{1}{2}\text{Tr}\left[\hat{\rho}(\delta q)\{\hat{\mathcal{L}}_{\delta q_i},\hat{\mathcal{L}}_{\delta q_j}\}\right], \quad i,j \in \{1,2\},
\end{equation}
where $\{, \}$ denotes the anticommutator and $\hat{\mathcal{L}}_{\delta q_i}$ is the symmetric logarithmic derivative (SLD) operator,
\begin{equation}\label{eq:SLDrelation}
\frac{\partial}{\partial\delta q_i}\hat{\rho}(\delta q)=\frac{1}{2}\left\{\hat{\rho}(\delta q), \hat{\mathcal{L}}_{\delta q_i}\right\}.
\end{equation}

We have already obtained the phase space representation $W(\delta q)$ of the density matrix $\hat{\rho}(\delta q)$ and therefore similarly to our approach 
in Ref.~\cite{SanavioBernadXuereb2020}, we derive the QFIM from the Gaussian Wigner function in Eq. \eqref{Wigner}. We neglect the subscripts of $\sigma^{\text{out}}_l$ in the subsequent discussion because we  
focus on the only mode of the electromagnetic field that is subject to detection, i.e.,  $\sigma=\sigma^{\text{out}}_l$. Furthermore, we also write $R=(X^{\text{out}}_l,Y^{\text{out}}_l)^T=(x,y)^T$.  

The Weyl transform of the $2$ SLD operators $\mathcal{L}_{\delta q_i}$ is quadratic 
and can be written as
\begin{equation}\label{eq:WeylSLD}
L^i(x,y)=R^T\Phi^i R-\nu^i,
\end{equation}
with $\Phi^i=-\partial_{\delta q_i}(\sigma^{-1})/2$ and $\nu^i=Tr[\Phi^i\sigma]$. Consequently, we find the Weyl transform $L^{(2)}_{ij}(x,y)$ of the operator 
$\frac{1}{2}\{\hat{\mathcal{L}}_{\delta q_i},\hat{\mathcal{L}}_{\delta q_j}\}$, see the details in Appendix \ref{app:WeylTransform}. Now, we can calculate the elements of QFIM by using the phase space representation
as
\begin{equation}
\mathcal{H}_{ij}=\int dx \, dy \, L^{(2)}_{ij}(x,y)W(x,y).
\end{equation} 
Finally, we obtain
\begin{eqnarray}\label{eq:QFIcomponentsWT}
\mathcal{H}_{ij}&=&3 \text{Tr}[(\Phi^i\sigma\Phi^j\sigma)]-\nu^i\nu^j\\
&&+\bigg{(}\det\sigma-\frac{1}{2}\bigg{)}(\Phi_{11}^i\Phi_{22}^j+\Phi_{11}^j\Phi_{22}^i-2\Phi_{12}^i\Phi_{12}^j)\nonumber.
\end{eqnarray}
It is worth mentioning that in the case of $i=j$ \eqref{eq:QFIcomponentsWT} reduces to Eq.~(37) of Ref.\cite{SanavioBernadXuereb2020}.

On the other hand, the CFIM $\mathcal{F}$ depends on the PDF of the measurements. The entries are
\begin{equation}\label{eq:CFIMatrix}
\mathcal{F}_{ij}=\int dk P(k; \delta q)\left[\partial_{\delta q_i} \ln P(k; \delta q) \right] \left[\partial_{\delta q_j}\ln P(k; \delta q)\right],
\end{equation}
where $i,j \in \{1,2\}$ and $P(k; \delta q)$ is the PDF parameterized by the unknown $\delta q$, which describes the probability of observing the outcome $k$. As we have already outlined, 
we consider BHD measurements, which has been proved in Ref. \cite{SanavioBernadXuereb2020} to be an optimal measurement for the inference of the optomechanical coupling 
strength in a standard moving-end mirror setup. The Weyl transform of the BHD POVM is
\begin{equation}\label{BHDPOVM}
\Pi_k^\eta(x,y)= \sqrt{\frac{2\eta}{\pi (1-\eta)}}\exp\bigg{[}-\frac{2\eta(k-\frac{x\cos\theta+y\sin\theta}{\sqrt{2}})^2}{1-\eta}\bigg{]}.
\end{equation}
where $k$ is an outcome of the measurement, $\eta$ is the detector efficiency and $\theta$ is the measured phase quadrature. This formula is usually obtained by considering an intense coherent local oscillator that
interferes with the single mode field state to be measured at a $50/50$ beam splitter. Then, the two modes emerging from the
beam splitter are measured by two photodetectors and the difference of the photon numbers $n_{12}$ is retained. These considerations yield $k=n_{12}/(2 \eta |\alpha_{LO}|)$  where $|\alpha_{LO}|^2$ is the mean 
photon number of the local oscillator's state.

The PDF $P(k; \delta q)$ is obtained by integrating the product of the phase space representation of BHD in Eq. \eqref{BHDPOVM}
and the Wigner function,
\begin{equation}\label{pdfk}
P(k; \delta q)=\sqrt{\frac{r_\theta^\eta(\sigma)}{2 \pi}}\exp\left[-r_\theta^\eta(\sigma)k^2/2\right],
\end{equation}
where we have introduced the function
\begin{eqnarray}
r_\theta^\eta(\sigma)=\frac{4\eta}{1-\eta+2\eta R^T_\theta\sigma R_\theta} \nonumber
\end{eqnarray}
with $R_\theta=(\cos\theta,\sin\theta)^T$. Now, we employ Eq.~\eqref{eq:CFIMatrix} to find the entries of matrix $\mathcal{F}$ and get  
\begin{equation}
\mathcal{F}_{ij}= \begin{cases}
2 \eta^2 \frac{\left(R_\theta^T\partial_{\delta q_i}\sigma R_\theta\right) \left( R_\theta^T\partial_{\delta q_j}\sigma R_\theta\right)}{\left(1-\eta+2\eta R_\theta^T\sigma R_\theta\right)^2}, & \mbox{if } i \ne j  \\
 2 \eta^2 \bigg{(}\frac{ R_\theta^T\partial_{\delta q_i}\sigma R_\theta}{1-\eta+2\eta R_\theta^T\sigma R_\theta}\bigg{)}^2, & \mbox{if } i=j \end{cases}
\end{equation}

To search for conditions under which the remoteness between CFIM and QFIM is as small as possible we employ the trace norm to quantity this distance by
\begin{equation}\label{distance}
 d=\|\mathcal{H}-\mathcal{F}\|_1.
\end{equation}
This norm distance $d$ is a function of all parameters of the model and the measurement scenario as well. In the multiparameter estimation scenarios usually, there is no optimal measurement to reach equality 
$\mathcal{H}=\mathcal{F}$ \cite{Matsumoto}. In addition, we are only dealing with the subspace of all possible POVMs and our strategy will be to find the minimum of $d$ within the 
experimentally available parameter space.

Finally, we are going to show how classical estimation is going to work on the obtained data. Based on the PDF in Eq. \eqref{pdfk} an experiment can obtain a finite sample 
${\bf k}=\{k_1,, k_2, \dots k_N\}$. After observing ${\bf k}$, we shall want to estimate the values of $\delta q$. We denote this estimate in vector notation as $\delta \tilde{q}({\bf k})$, 
which is the estimator applied on the data space. We assume that all observations are effectively independent because the values of the integrated photocurrents in BHD are recorded per pulse \cite{Raymer}.
Then,
\begin{equation}
 P({\bf k}; \delta q)=\prod^N_{i=1} \sqrt{\frac{r_\theta^\eta(\sigma)}{2 \pi}}\exp\left[-r_\theta^\eta(\sigma)k^2_i/2\right].
\end{equation}
It is straightforward to check that
\begin{equation}
 \int d{\bf k}\, P({\bf k}; \delta q)\left[\partial_{\delta q_i} \ln P({\bf k}; \delta q) \right]=0, \quad \forall \delta q_i
\end{equation}
with $i \in \{1,2\}$. Therefore, an unbiased estimator $\delta \tilde{q}({\bf k})$ attains the Cram\'er-Rao lower bound if and only if \cite{vanTrees}
\begin{equation}\label{attaining}
 \frac{\partial \ln P({\bf k}; \delta q) }{\partial \delta q}=\mathcal{I} (\delta q) \left[\delta \tilde{q}({\bf k})- \delta q  \right],
\end{equation}
where $\mathcal{I}$ is some $2 \times 2$ matrix. The left-hand side of \eqref{attaining} reads
\begin{eqnarray}
 &&\frac{\partial \ln P({\bf k}; \delta q) }{\partial \delta q}=\begin{bmatrix}\frac{\partial \ln P({\bf k}; \delta q) }{\partial \delta q_1} \\ 
 \frac{\partial \ln P({\bf k}; \delta q) }{\partial \delta q_2} \end{bmatrix}  \\
 &&=\begin{bmatrix} \frac{\eta R_\theta^T\partial_{\delta q_1}\sigma R_\theta}{1-\eta+2\eta R_\theta^T\sigma R_\theta}\left(\frac{4 \eta \sum^N_{i=1}k^2_i }{1-\eta+2\eta R_\theta^T\sigma R_\theta}-N \right)   \\ 
 \frac{\eta R_\theta^T\partial_{\delta q_2}\sigma R_\theta}{1-\eta+2\eta R_\theta^T\sigma R_\theta}\left(\frac{4 \eta \sum^N_{i=1}k^2_i}{1-\eta+2\eta R_\theta^T\sigma R_\theta}-N \right)  \end{bmatrix}. \nonumber
\end{eqnarray}
We observe that this vector cannot be written in the form required by \eqref{attaining}
\begin{equation}
 \begin{bmatrix} \mathcal{I}_{11} (\delta q) & \mathcal{I}_{12} (\delta q) \\ \mathcal{I}_{21} (\delta q) & \mathcal{I}_{22} (\delta q)
  \end{bmatrix} \begin{bmatrix}\delta \tilde{q}_1({\bf k})- \delta q_1 \\ 
 \delta \tilde{q}_2({\bf k})- \delta q_2 \end{bmatrix},
\end{equation}
and therefore, an efficient unbiased estimator does not exist. However, one can still look for minimum variance unbiased estimators by using the concept of complete sufficient statistics and the 
Rao-Blackwell-Lehmann-Scheffe theorem \cite{Kay, Casella}. By examining the PDF one can realize that
\begin{equation}
 T({\bf k})= \left[
                   \sum^N_{i=i} k^2_i, \quad \sum^N_{i=i} k^2_i \right]^T.
\end{equation}
is a sufficient statistic for $\delta q_1$ and $\delta q_2$. Taking the expectation value produces
\begin{equation}
   \int d{\bf k} \, P({\bf k}; \delta q) \,T({\bf k})=\begin{bmatrix}
                                                               N \frac{1-\eta+2\eta R^T_\theta\sigma R_\theta} {4\eta} \\ N \frac{1-\eta+2\eta R^T_\theta\sigma R_\theta} {4\eta}
                                                              \end{bmatrix}.
\end{equation}
The task is to find two functions $f_1$ and $f_2$ such that
\begin{equation}
 \int d{\bf k} \, P({\bf k}; \delta q) f_i\left[T({\bf k})\right]=\delta q_i, \quad i\in\{1,2\}. 
\end{equation}
However, this turns out to be difficult because $\sigma$ is the solution of the Lyapunov equation \eqref{eq:Lyapunov} where $A$ contains $\delta q_1$ and $\delta q_2$. In fact, $\sigma$ is a function of the 
eigenvalues and eigenvectors 
of $A$, which depend on the parameters to be estimated. As $A$ is either a $6$ or $10$ dimensional matrix in our models (TR and CC) and the fact that finding analytical roots to general polynomial equations 
of degree five or higher is not possible shows that the two functions $f_1$ and $f_2$ cannot be determined analytically.   

We have seen so far that the above two attempts fail analytically and the complete sufficient statistic approach may work with a considerable numerical effort. 
As next, the maximum likelihood approach could be tried, if $P({\bf k}; \delta q)$, the likelihood function, 
can be maximized either analytically or numerically. The likelihood equations are 
\begin{equation}
 \left. \frac{\partial \ln P({\bf k}; \delta q) }{\partial \delta q_i}\right |_{\delta q=\delta\tilde q({\bf k})}=0, \quad i\in\{1,2\},
\end{equation}
which yield two equations. These equations differ only in a non-zero factor and because on the right-hand side stays zero, we get only one equation to be solved
\begin{equation}
 R_\theta^T\sigma \Big(\delta \tilde{q}_1({\bf k}),\delta\tilde{q}_2({\bf k}) \Big) R_\theta=\frac{2}{N} \sum^N_{i=i} k^2_i-\frac{1-\eta}{2\eta}.
\end{equation}
This equation has to be solved numerically for a given sample ${\bf k}$ including the second partial derivative test with the Hessian matrix of $P({\bf k}; \delta q)$, which assures that we have found 
the maximum of the likelihood function. This approach guarantees estimates which are efficient asymptotically, i.e., $N \to \infty$. If one cannot succeed with the maximum likelihood 
approach then there is still the method of moments, however, these estimators are not optimal, and extracting the estimates of $\delta q_1$ and $\delta q_2$ out of 
$\sigma$ can only be solved numerically. 

\section{Results}
\label{sec:IV}

In this section, we numerically investigate the norm distance $d$ between CFIM and QFIM for an experimentally feasible situation. In Sec. \ref{sec:III} we have discussed the strategy of the estimation and argued
that the estimators of the disorders can be found numerically from the measurement data. Therefore, we aim to minimize $d$ for the experimentally tunable parameters so that the postprocessing of the measurement 
data results in estimators with variance close to the benchmark value defined by QFIM. We are going to analyze both the CC and TR models presented in Sec. \ref{sec:II}. 

For our numerical analysis, we take the experimental 
values from \cite{Piergentili2018}, where the optomechanical interaction has been studied for different input powers of the driving field. The cavity intensity decay rate $\kappa/2 \pi$ was found to be $83$ kHz. We
consider the CC model to possess equal decay rates $\kappa_1=\kappa_3=\kappa$ for the first and third inner cavities. Furthermore, we also assume that the middle inner cavity decay rate $\kappa_2 \ll \kappa$,
because the two membranes may absorb photons or scatter them out of the cavity, but this loss is negligible compared to the photon leakage at the end mirrors. In the TR model, there is only one decay rate. 
Photodetectors are considered to stay on for a temporal window of length $\tau=1/\kappa$. Intensity $\varepsilon$ of the driving field is equal to $\sqrt{2\kappa P/\hbar \omega_{\text{L}}}$, where $P$ is the power 
of the laser. The largest optomechanical coupling strength $g/2 \pi= 0.30$ Hz was obtained for low power, i.e, $P=130$ $\mu$W, with $\gamma/2 \pi=1.64$ Hz and $\omega_{\text{m}}/2 \pi=235.81$ MHz. Both membranes have the same masses 
$m=0.72$ ng and reflectivities $r=0.33$, while the experiment was performed at room temperature $T=300$ K. The low reflectivity of the membranes indicates that this experiment corresponds more to the TR model.

In order to address the CC model as well, we need to assume that the same experiment can be carried out with different membranes yielding much larger reflectivity values. In this context, 
the hopping rate $J$ which couples the modes of the CC model is obtained by setting 
the three inner cavity mode amplitudes to be approximately the same. An application of a driving laser from the left populates the mode of the left inner cavity, and without a sufficient large hopping rate there
is a risk of leaving the mode of the right inner cavity very low populated or empty, and thus making impossible the detection procedure. 
A proper choice of the hopping constant, in our case a value $J\approx 200$kHz, prevents this to happen. With the formula $J=\omega_c \sqrt{2(1-r)}$ \cite{Jayich2008}, we can find the reflectivity of our membrane, 
yielding $r\approx 1$, which makes the CC model suitable to describe the system.

Our aim is to investigate the CFIM and the QFIM around these experimental values. 
The CFIM depends also on the detectors efficiency $\eta$ and the phase $\theta$ of the BHD. We assume $\eta=1$, as existing detectors are already close to ideals \cite{Daiss} and the destructive effects
of non-ideal detection efficiency are known \cite{SanavioBernadXuereb2020}. Taking the inverse of CFIM and investigating the diagonal elements, which are the lower bounds of the variances of the estimators 
$\delta \tilde{q}_1$ and $\delta \tilde{q}_2$ in this BHD scenario, one can understand the dependence on the phase $\theta$. 
We have retrieved minimum values at $\tilde{\theta}^{(\text{CC})}=0$, and $\tilde{\theta}^{(\text{TR})}=\pi/2$, both with a period of $\pi$. Those values can be obtained numerically and depend strongly on the 
experimental values considered.

\begin{figure}
\includegraphics[width=.42\textwidth]{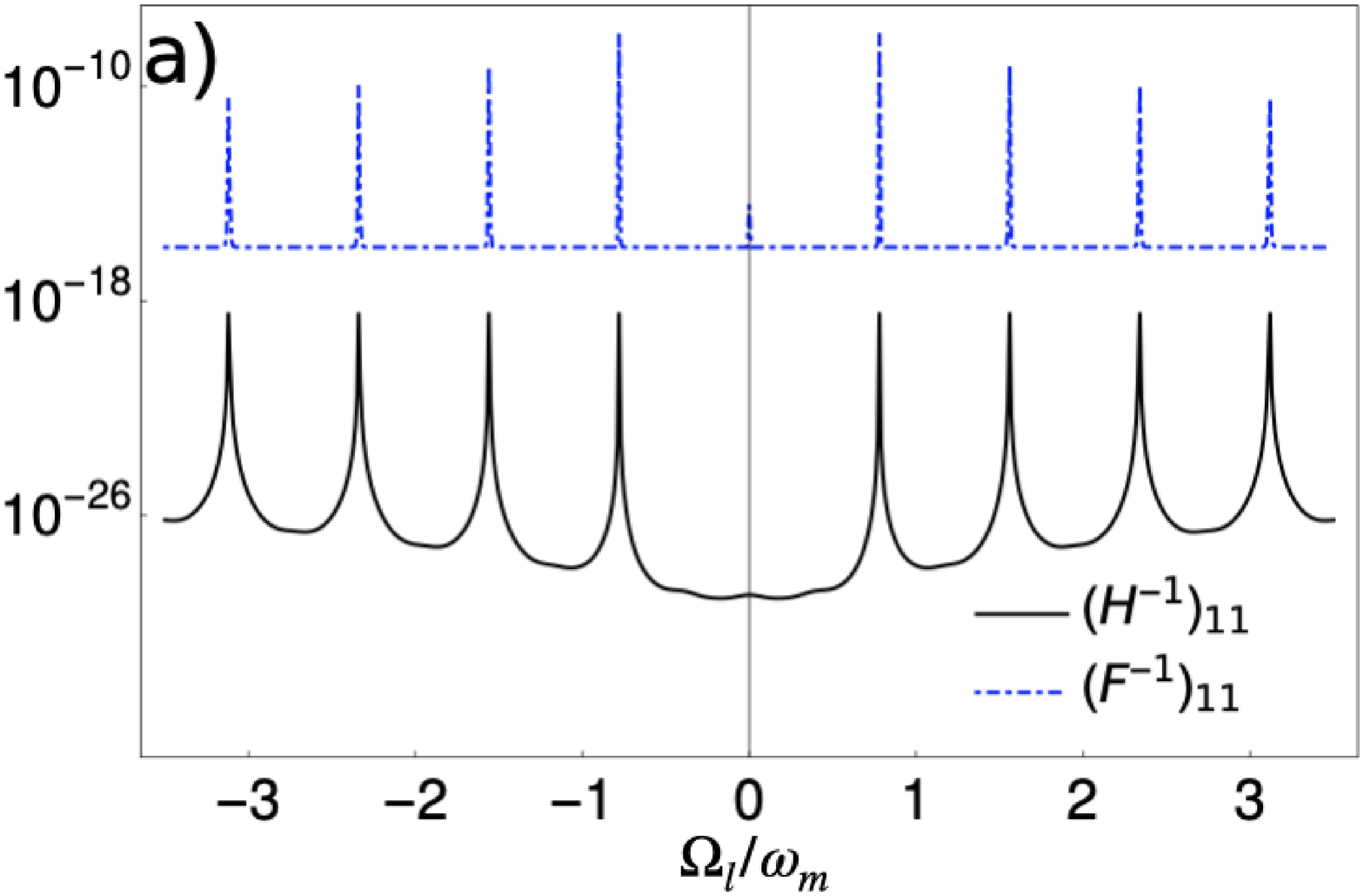}
\includegraphics[width=.42\textwidth]{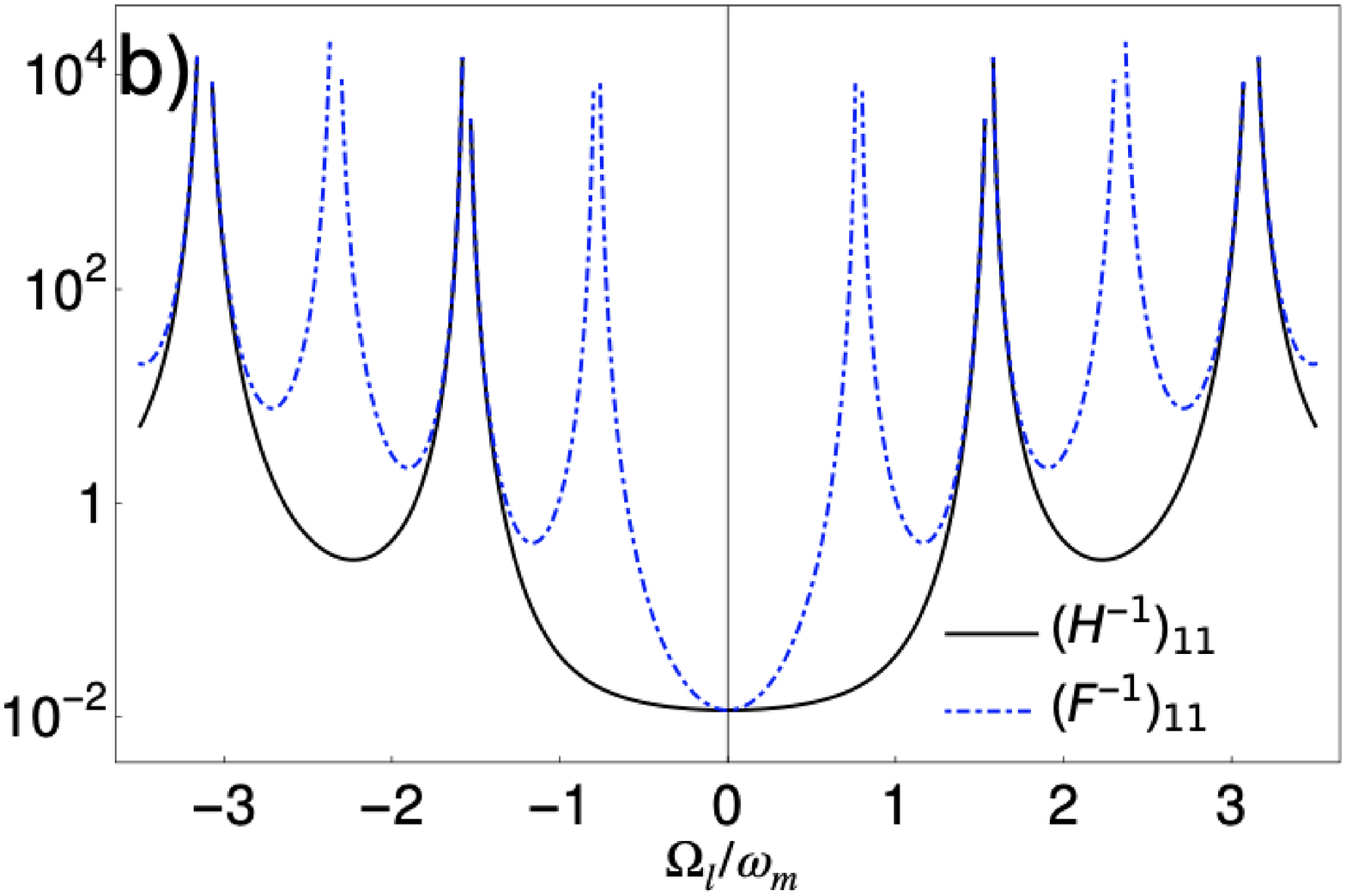}
\caption{Semi-logarithmic plot of quantum and classical lower bounds of the variance $Var(\delta q_1)$ expressed in $\text{m}^2$ as a function of $\Omega_l/\omega_m$. a) CC model. b) TR model. 
The experimental values are taken from \cite{Piergentili2018}.}
\label{fig:VarvsOmega}
\end{figure}

Once we have optimized for the detector's phase, we need to understand which central frequency $\Omega_l$ of the filter function gives us the best accuracy on the estimation of the disorders. 
Therefore, one has to calculate the inverse of QFIM and investigate both diagonal elements of $H^{-1}$, which are the smallest lower bounds of the variances of the estimators $\delta \tilde{q}_1$ and 
$\delta \tilde{q}_2$.
Fig. \ref{fig:VarvsOmega} shows that the minimum variance is obtained at $\Omega_l=0$, i.e. in correspondence of the frequency of the driving laser. This result is valid for both the CC and the TR models. 

Beside this similarity, the two models don't share the same features. In fact, whereas for the TR model the BHD appears to be an optimal measurement, as the classical and quantum lower bounds 
for the variance coincide, 
for the CC model this measurement scenario is far from saturating inequality \eqref{eq:inequality} as both the diagonal components of the inverse of classical and quantum Fisher information matrix differ of 
many orders of magnitude. 
However, it is worth to notice that when the CC model is considered, BHD is able to offer estimates of disorders in the positions of the two membranes with extreme accuracy, i.e.,
 $Var(\delta\tilde{q}_1)$ and $Var(\delta\tilde{q}_2)$ $\sim10^{-10}-10^{-16}\, \text{m}^2$.

\begin{center}
\begin{figure}
\includegraphics[width=.42\textwidth]{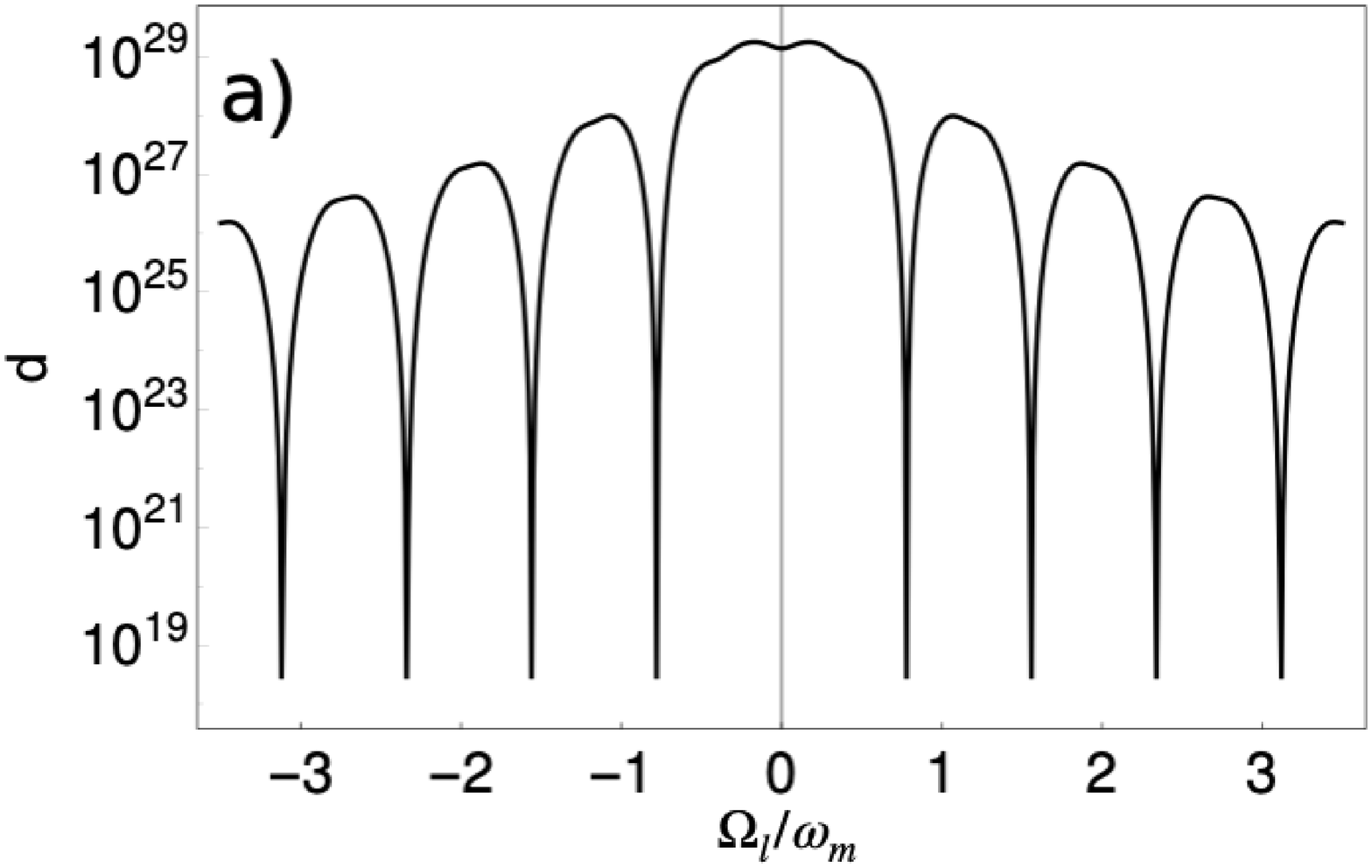}
\includegraphics[width=.42\textwidth]{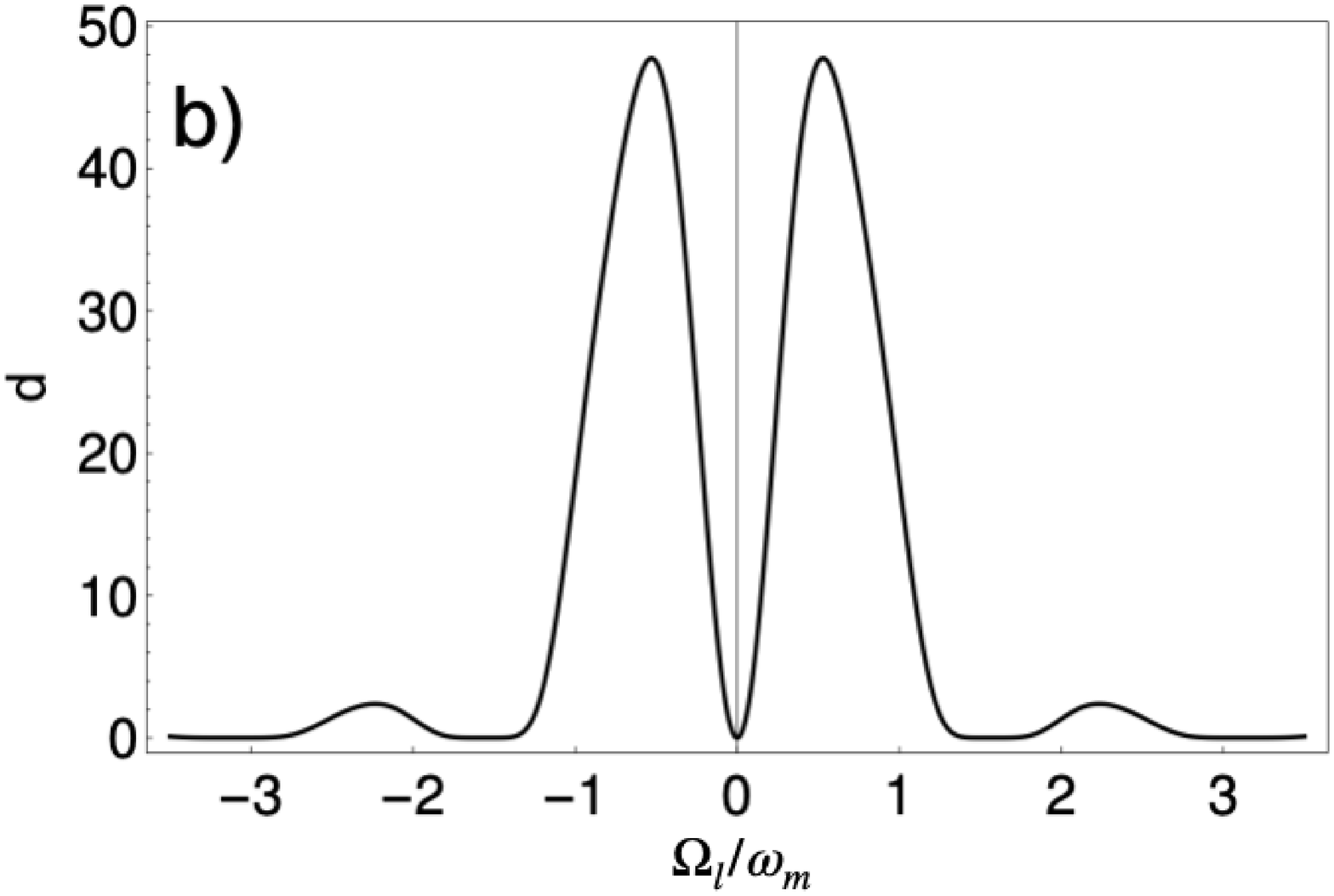}
\caption{Semi-logarithmic plot of distance $d$ in trace norm as a function of $\Omega_l/\omega_m$. a) CC model. b) TR model. The experimental values are taken from \cite{Piergentili2018}.}
\label{fig:dvsOmega}
\end{figure}
\end{center}

Fig.~\ref{fig:dvsOmega} shows the distance $d$ in trace norm as a function of the filter frequency $\Omega_l$. For the CC model, the CFIM and the QFIM are far from each other, as $d$ is very large, 
suggesting the BHD is not the optimal measurement. However, we can be relieved by the fact the variances at $\Omega_l=0$ are very small (see Fig.~\ref{fig:VarvsOmega}).
This is different in the TR model, where under optimal conditions ($\eta=1,\theta=\tilde{\theta}^{(\text{TR})}$), we have found that $d$ goes to zero when $\Omega_l=0$. We notice that $d$ is very small also 
for other values of $\Omega_l$, but on those points, the lower bound of the variance is larger (see Fig.\ref{fig:VarvsOmega}). This results in a poor estimation of the membrane position, 
with an uncertainty larger than the size of the cavity itself. This condition can easily be overcome by taking enough number $N$ of identical and independent measurements, which ultimately decreases the 
lower bound by a factor of $1/N$. Our analysis shows that in the TR model little information about the position of the membranes is contained in the state of the output field.
Therefore, one has to tune the system parameters such that $d$ and the lower bound of the variance are getting close to a minimum.

\begin{figure}
\includegraphics[width=.45\textwidth]{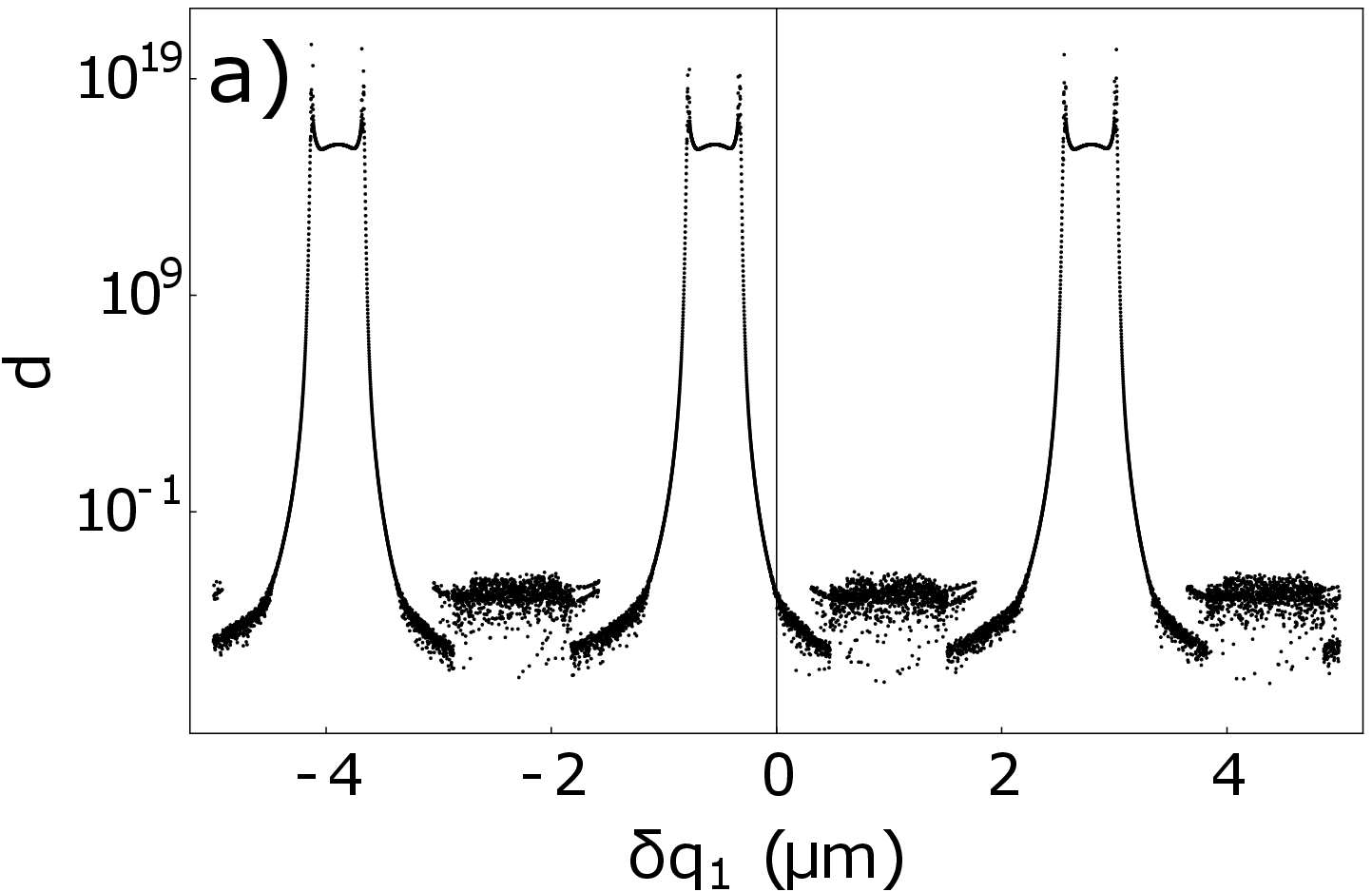}
\includegraphics[width=.43\textwidth]{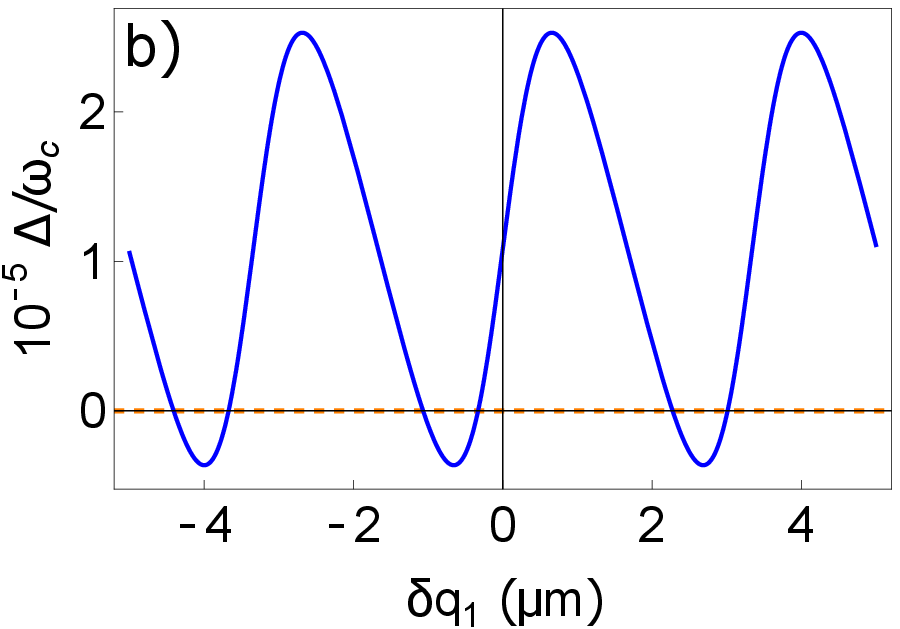}
\caption{a) Semi-logarithmic plots of distance $d$ in trace norm as a function of the disorder $\delta{q}_1$. b) The cavity frequency obtained by varying the prepared position of the first membrane with the 
disorder $\delta{q}_1$ and keeping $\delta{q}_2=0$. 
In b) the horizontal line refers to the cavity natural frequency in absence of membranes. Both figures belong to the TR model.}
\label{fig:dvsdq1}
\end{figure}

\begin{figure}
\includegraphics[width=.44\textwidth]{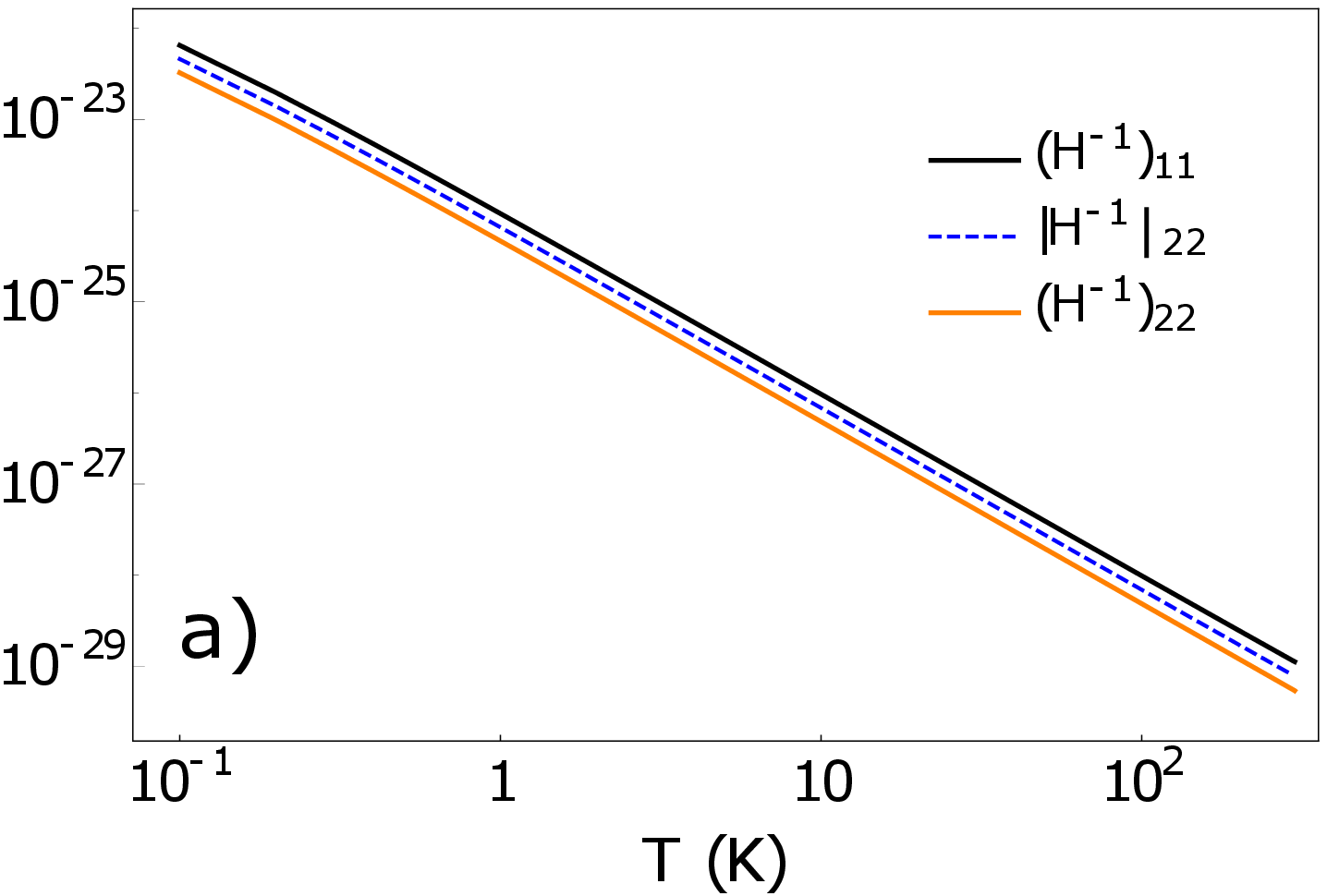}
\includegraphics[width=.44\textwidth]{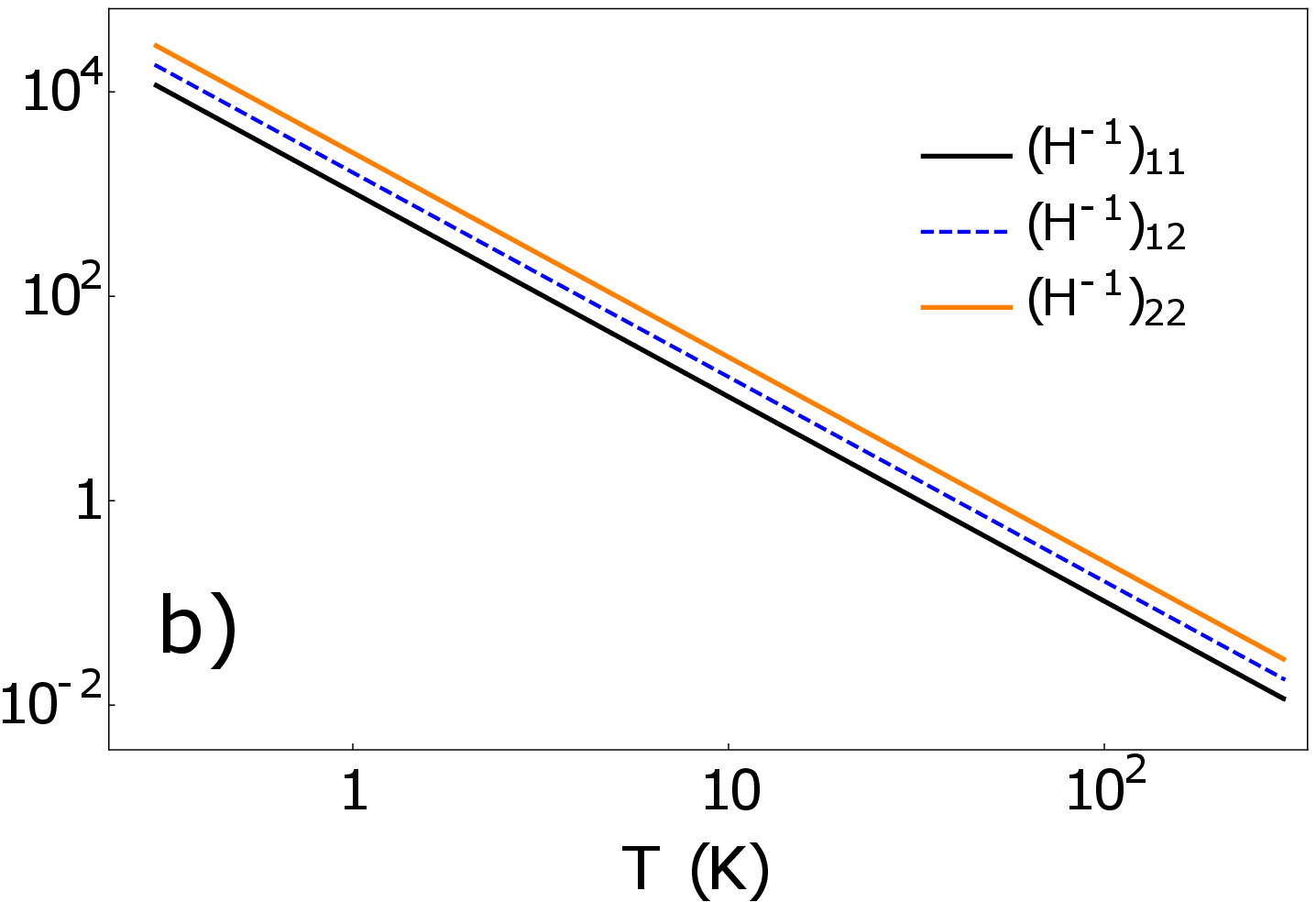}
\caption{Log-log plots of the quantum lower bounds of the variances expressed in $\text{m}^2$ as a function of the temperature $T$. a) CC model. b) TR model. $H^{-1}_{12}$ gives information on the correlation
of the data used for estimating the two disorders $\delta{q}_1$ and $\delta{q}_2$.}
\label{fig:HvsT}
\end{figure}

The true values of the disorders largely modify the value of $d$. In Fig.~\ref{fig:dvsdq1} we plot the resulting distance $d$ in trace norm for the TR model, calculated as a function of 
$\delta{q}_1$, keeping $\delta{q}_2=0$. For a possible value of the disorder $\delta{q}_1\sim0.5\mu$m, the distance between CFIM and QFIM is further reduced and the estimation gets closer to the optimal. 
Analogous results are obtained when we keep $\delta{q}_1$ fixed and we vary the disorder for the other membrane. We notice that $d$ has the 
same period of the cavity frequency as expressed in Eq.~\eqref{eq:piergentiliformula}, and its minima are reached when $\omega_c$ is at maximum.
Fig. \ref{fig:HvsT} shows how the variances lower bounds decreases with increasing temperature. The reason for this unexpected result has to be searched in the noise matrix $D$ of 
Eq. \eqref{eq:DMatrix}, from which we derive the correlation matrix $\sigma$, that has terms proportional to $T$. The off-diagonal component of the inverse matrix $\mathcal{H}^{-1}$ decreases, as 
the increase of temperature lowers the correlations between the two membranes.

\begin{figure}
\includegraphics[width=.42\textwidth]{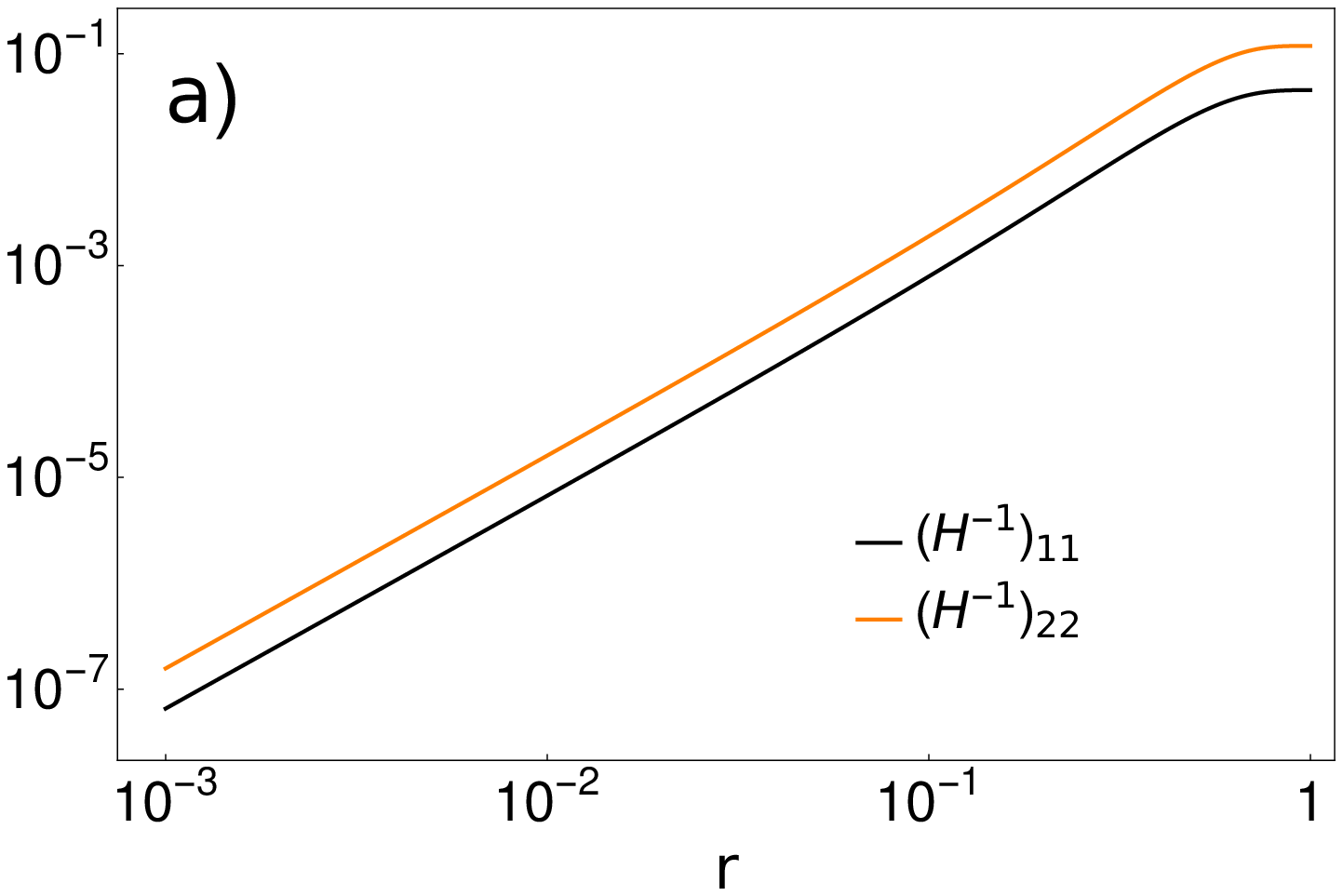}
\includegraphics[width=.42\textwidth]{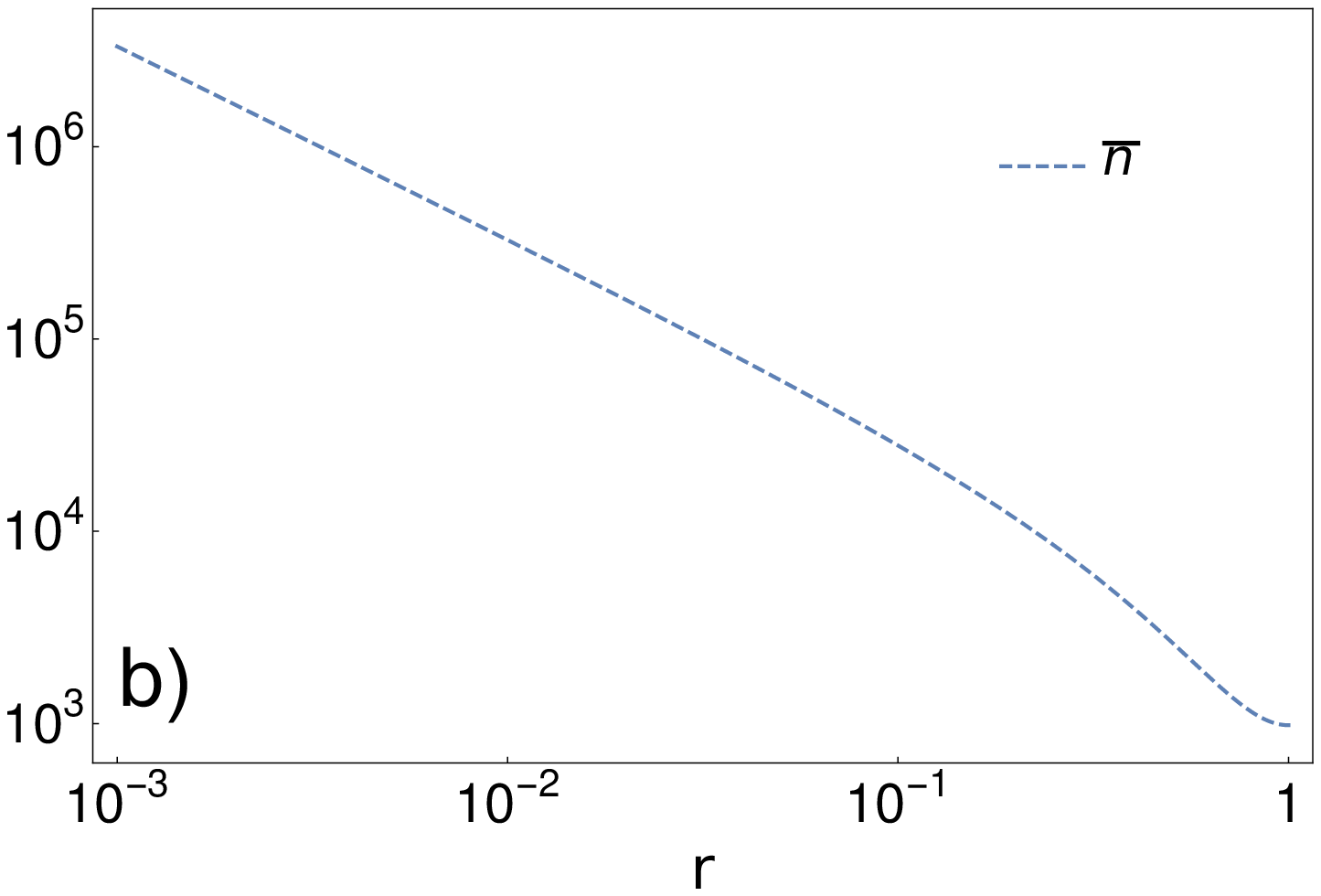}
\caption{a) Log-log plot of the quantum lower bounds of the variances in $\text{m}^2$ as a function of the reflectivity $r$. b) Log-log plot of the mean photon number $\bar{n}$ in the cavity as a function 
of the reflectivity $r$. Both figures belong to the TR model.}
\label{fig:TRtoCC}
\end{figure}

Finally, we consider only the TR model and we check the results when we change the reflectivity of the two membranes. Whereas the CC model is defined only for a high reflectivity $r\approx 1$ membrane, 
the TR model can be used for any value of $r$. Fig. \ref{fig:TRtoCC} shows the lower bounds of the variance increases with the reflectivity $r$ 
of the membrane, where we found it varies as the inverse of $\bar{n}^2=|\alpha|^4$, squared mean photon number in the cavity. 
The high reflectivity screens the radiation from passing through the membranes and lowers the rate of photons leaving the cavity, which results in an increased lower bound of the variance.

\section{Conclusions and Discussions}
\label{sec:V}

In this paper, we have investigated an optomechanical setup with a driven cavity containing two oscillating membranes. We have considered two possible theoretical models for the description of this system. The CC 
model focuses on a case, where three coupled single modes of the electromagnetic field are present in the inner cavities defined by the two membranes and the mirrors of the cavity. In the TR model, it is assumed 
that a global single mode of the radiation field is present in the whole cavity. Range of applicability of these models strongly depends on the reflectivity of the membranes. Our models consider also 
high-temperature quantum Brownian motions of the membranes, photon losses of the cavity fields, and the input-output formalism for the description of the output field escaping the cavity. In typical cavity optomechanical 
experiments, the estimation of parameters like the optomechanical coupling is done by detecting the light transmitted by the cavity, which is similar to our theoretical approach presented here. 
Within the CC and TR models, we have considered estimations of disorders in the positions of the membranes. For these estimations, the data is obtained via BHD of the escaping field and thus the estimators of 
disorders mapping the data into estimates have accuracies related to the CFIM. The quantum optimal accuracies are obtained from the QFIM. Without solving, in particular, the attainability of the CFIM related 
bounds by some unbiased estimators, we have focused from a purely theoretical point of view on the attainability of QFIM by CFIM. It is indeed true that most of the estimators even during classical postprocessing 
of data are unable to attain the Cram\'er-Rao bound \cite{Kay}, but there is still a well-understood decision-making process in estimator selection. In this view, our analysis serves the purpose of characterizing 
the chosen measurement setup for a certain estimation case, here the estimation of the disorders.

A comparison of CFIM and QFIM shows that the phase $\theta$ of the local oscillator in the BHD results in specific angles for an optimal estimation. In the unit detector efficiency limit we have obtained 
$\theta=0$, i.e, measuring the distribution of quadrature $X$ of the output field, for the CC and $\theta=\pi/2$, i.e, measuring the distribution of quadrature $P$ of the output field, for the TR model. 
This marked contrast could be an important help for experimental setups with different reflectivities of the membranes. With respect to the frequency of the filtering function used in input-output relations, 
both models deliver different optimal frequencies for the distance of QFIM and CFIM. However, it still seems when the frequency of the filter function matches the frequency of the driving laser a good enough 
accuracy can be obtained. Actually, there is an interesting effect, namely for certain values of parameters the CFIM might saturate QFIM, however, the related accuracies of the estimation could be very bad. 
Whenever is this the case we have indicated it, because we need not only to obtain saturation, but we have also to make sure that the related estimation precisions are good enough. This applies also to the 
effects of temperature, where the distance between CFIM and QFIM is increasing with the increase of the temperature, predictable behavior of the system. However, the accuracies of the estimators are getting 
better with the increase in temperature. This means that warmer baths of the membranes result in better precisions, a similar effect found by Ref. \cite{Sala}, but on the other hand, reaching the quantum optimal 
limit becomes more and more distant.

We have seen very different results whenever the considered model is the CC or the TR model. The two scenarios have shown lower bounds of variances of estimators with very different scales. 
Our choice for the hopping rate value $J$ has led the mean photon numbers in the three cavities of the CC model to be very different from the optical amplitude of the delocalized mode of the TR model.  Furthermore, the two models are characterized by different expressions for the cavity frequency and consequently for the optomechanical coupling strength. 
It's worth to notice that the CC model offers a better description of the physics of some optomechanical lattice systems \cite{Schmidt,SanavioPeano}, and the TR model is more suited when one optical mode is coupled to multiple membranes.
We believe that our implementable 
theoretical approach may serve the aim of realizing enhanced optomechanical performances \cite{Matheny,Li,Weaver}, the main 
objective of current experimental efforts.

Given the models considered here or in our previous work \cite{SanavioBernadXuereb2020} we can conclude that the probability density function of the data is always Gaussian, whose variance depends only on
the parameters to be estimated, but unfortunately in a complicated matter. In this paper we have described several approaches, which suggest numerical approaches for finding minimum variance unbiased estimators.
Therefore, a future goal may be to address this estimator selection issue analytically for this family of probability density functions.

\section{Acknowledgments}

This work is supported by the European Union's Horizon 2020 for research and innovation programme under Grant Agreement No.\ 732894 (FET Proactive HOT) and by the 
DFG under Germany's Excellence Strategy -- Cluster of Excellence Matter and Light for Quantum Computing (ML4Q) EXC 2004/1 -- 390534769. 
CMS is funded by the International Foundation of Big Data and Artificial Intelligence for Human Development
within the project “Quantum computing for data analysis”
\bibliographystyle{apsrev4-1}
\bibliography{biblio}

\begin{thebibliography}{48}
\bibitem [{com()}]{comment}%
  \BibitemOpen
  \href@noop {} {}It is important to note that we have misprints in Eqs. (25)
  and (27), see \cite{SanavioBernadXuereb2020}. Here, we correct
  them.\BibitemShut {Stop}%
\end{thebibliography}

\appendix
\begin{widetext}
 \section{Steady-state amplitudes and the dynamical matrix in the CC model}
\label{app:steadystateamplitude}

In the Heisenberg picture, the Hamiltonian in \eqref{eq:CCHamiltonian} together with the dissipative dynamics explained in \ref{sec:IIc} yield 
\begin{eqnarray}
\dot{\hat{a}}_1&=&-i\Delta_0 \hat{a}_1-ig\hat{a}_1\hat{q}_1+\varepsilon-iJ\hat{a}_2-\frac{\kappa_1}{2}\hat{a}_1+\sqrt{\kappa_1}\hat{a}_{\text{in},1},\\
\dot{\hat{a}}_2&=&-i\Delta_0 \hat{a}_2-ig\hat{a}_2 (\hat{q}_2-\hat{q}_1)-iJ\left(\hat{a}_1+\hat{a}_3\right)-\frac{\kappa_2}{2}\hat{a}_2+\sqrt{\kappa_2}\hat{a}_{\text{in},2},\\
\dot{\hat{a}}_3&=&-i\Delta_0 \hat{a}_3+ig\hat{a}_3\hat{q}_2-iJ\hat{a}_2-\frac{\kappa_3}{2}\hat{a}_3+\sqrt{\kappa_3}\hat{a}_{\text{in},3},\\
\dot{\hat{p}}_1&=&-m\omega_{\text{m}}^2\hat{q}_1-\gamma\hat{p}_1-\hbar g \left(\hat{a}^\dagger_1\hat{a}_1-\hat{a}^\dagger_2\hat{a}_2\right)+\hat{\xi}, \\
\dot{\hat{p}}_2&=&-m\omega_{\text{m}}^2\hat{q}_2-\gamma\hat{p}_2-\hbar g \left(\hat{a}^\dagger_2\hat{a}_2-\hat{a}^\dagger_3\hat{a}_3\right)+\hat{\xi}, \\
\dot{\hat{q}}_1&=&\frac{\hat{p}_1}{m}, \quad \dot{\hat{q}}_2=\frac{\hat{p}_2}{m},
\end{eqnarray}
and the dynamics of the hermitian conjugates of $\hat{a}_1$, $\hat{a}_2$, and $\hat{a}_3$. We introduce the following transformations $\hat{p}_i\to p_{i}+\hat{p}_i$, $\hat{q}_i\to q_{i}+\hat{q}_i$, and 
$\hat{a}_j\to\alpha_j+\hat{a}_j$, which can also be viewed as an application of different displacement operators to the master equation. In this case, one has to consider the dissipation of the 
field modes to be governed by the optical master equation~\cite{BreuerPetruccioni2002}, whereas the membranes follow the Caldeira-Leggett master equation~\cite{CaldeiraLeggett1981}.
In the steady-state, we obtain the following system of equations
\begin{eqnarray}
0&=&-i\Delta_0 \alpha_1-ig\alpha_1 q_1+\varepsilon-iJ\alpha_2-\frac{\kappa_1}{2}\alpha_1,\\
0&=&-i\Delta_0 \alpha_2-ig\alpha_2 (q_2-q_1)-iJ\left(\alpha_1+\alpha_3\right)-\frac{\kappa_2}{2}\alpha_2,\\
0&=&-i\Delta_0 \alpha_3+ig\alpha_3 q_2-iJ\alpha_2-\frac{\kappa_3}{2}\alpha_3,\\
0&=&-m\omega_{\text{m}}^2 q_1-\gamma p_1-\hbar g \left(|\alpha_1|^2-|\alpha_2|^2\right), \\
0&=&-m\omega_{\text{m}}^2 q_2-\gamma p_2-\hbar g \left(|\alpha_2|^2-|\alpha_3|^2\right), \\
0&=&\frac{p_1}{m}, \quad 0=\frac{p_2}{m}.
\end{eqnarray}
It is immediate that
\begin{eqnarray}
 q_1&=&\frac{\hbar g}{m\omega_m^2}\left(|\alpha_2|^2-|\alpha_1|^2\right), \quad p_1=0, \\
 q_2&=&\frac{\hbar g}{m\omega_m^2}\left(|\alpha_3|^2-|\alpha_2|^2\right), \quad p_2=0.
\end{eqnarray}
We can only find numerical solutions for the amplitudes $\alpha_1$, $\alpha_2$, and $\alpha_3$. Next, we introduce the quadratures 
$\hat X=(\hat a^\dagger + \hat a)/\sqrt{2}$ and $\hat Y=i(\hat a^\dagger - \hat a)/\sqrt{2}$ of the field operators. Then, we have the dynamical matrix
\begin{equation}
 A^{(CC)}=\begin{pmatrix}
                A_{11} & A_{12} \\
                A_{21} & A_{22}
               \end{pmatrix},
\end{equation}
acting on the vector of operators $(\hat{X}_1,\hat{Y}_1, \hat{X}_2, \hat{Y}_2,\hat{X}_3,\hat{Y}_3,\hat{p}_1,\hat{p}_2,\hat{q}_1,\hat{q}_2)^T$ with
\begin{eqnarray}
 A_{11}&=& \begin{pmatrix}
          -\frac{\kappa_1}{2} & \Delta_1 & 0 & J & 0 & 0 \\
          -\Delta_1 & -\frac{\kappa_1}{2} & -J & 0 & 0 & 0 \\
           0 & J & -\frac{\kappa_2}{2} & \Delta_2 & 0 & J \\
           -J & 0 & -\Delta_2 & -\frac{\kappa_2}{2} & -J & 0 \\
           0 & 0 & 0 & J & -\frac{\kappa_3}{2} & \Delta_3 \\
           0 & 0 & -J & 0 & -\Delta_3 & -\frac{\kappa_3}{2} 
         \end{pmatrix}, \\
 \Delta_1&=&\Delta_0+g q_1, \quad \Delta_2=\Delta_0+g \left(q_2-q_1\right), \quad \Delta_3=\Delta_0-g q_2, \\        
 A_{12}&=& \begin{pmatrix}
            0 & 0 & \sqrt{2}g\text{Im}[\alpha_1] & 0 \\
            0 & 0 & -\sqrt{2}g\text{Re}[\alpha_1] & 0 \\
            0 & 0 & -\sqrt{2}g\text{Im}[\alpha_2] & \sqrt{2}g\text{Im}[\alpha_2] \\
            0 & 0 & \sqrt{2}g\text{Re}[\alpha_2] & -\sqrt{2}g\text{Re}[\alpha_2] \\
            0 & 0 & 0 & -\sqrt{2}g\text{Im}[\alpha_3] \\
            0 & 0 & 0 & \sqrt{2}g\text{Re}[\alpha_3]
           \end{pmatrix}, \\
 A_{21}&=& \begin{pmatrix}
            -\sqrt{2}\hbar g\text{Re}[\alpha_1] & -\sqrt{2}\hbar g\text{Im}[\alpha_1] & \sqrt{2}\hbar g\text{Re}[\alpha_2] & \sqrt{2}\hbar g\text{Im}[\alpha_2] & 0 & 0 \\
            0 & 0 & -\sqrt{2}\hbar g\text{Re}[\alpha_2] & -\sqrt{2}\hbar g\text{Im}[\alpha_2] & \sqrt{2}\hbar g\text{Re}[\alpha_3] & \sqrt{2}\hbar g\text{Im}[\alpha_3] \\
            0 & 0 & 0 & 0 & 0 & 0\\
            0 & 0 & 0 & 0 & 0 & 0
           \end{pmatrix}, \\
A_{22}&=& \begin{pmatrix}
           -\gamma  & 0 & -m \omega_m^2 & 0 \\
           0& -\gamma  & 0 & -m \omega_m^2 \\
           \frac{1}{m} & 0 & 0 & 0 \\
           0& \frac{1}{m} & 0 & 0 
          \end{pmatrix}.           
\end{eqnarray}
One has to analyze the stability of the dynamical matrix, checking that each eigenvalue of $A^{(CC)}$ has a negative real part. This condition is necessary to express the steady-state as a 
Gaussian state. In our numerical simulations this condition is always satisfied.

\section{Steady-state amplitudes and the dynamical matrix in the TR model}
\label{app:BCCmodel}

The linearization of the dynamics involving Hamiltonian in \eqref{eq:HamiltonianOneMode} follows the same principles we saw for the CC model in Appendix \ref{app:steadystateamplitude}. 
Nevertheless, the corresponding equations are different as only one mode interacts with the mechanical oscillation of the membranes. 
In the Heisenberg picture, the resulting differential equations are
\begin{eqnarray}
\dot{\hat{a}}&=&-i\Delta_0 \hat{a}-i\hat{a}\left(g_1\hat{q}_1+g_2\hat{q}_2\right)+\varepsilon-\frac{\kappa}{2}\hat{a}+\sqrt{\kappa}\hat{a}_{\text{in}},\\
\dot{\hat{p}}_1&=&-m\omega_{\text{m}}^2\hat{q}_1-\gamma\hat{p}_1-\hbar g_1 \hat{a}^\dagger \hat{a}+\hat{\xi}, \\
\dot{\hat{p}}_2&=&-m\omega_{\text{m}}^2\hat{q}_2-\gamma\hat{p}_2-\hbar g_2 \hat{a}^\dagger \hat{a}+\hat{\xi}, \\
\dot{\hat{q}}_1&=&\frac{\hat{p}_1}{m}, \quad \dot{\hat{q}}_2=\frac{\hat{p}_2}{m},
\end{eqnarray}
and the dynamics of the hermitian conjugates of $\hat{a}$. In the steady-state, after performing the transformations shown in Appendix \ref{app:steadystateamplitude} we obtain the following system of equations
\begin{eqnarray}
0&=&-i\Delta_0 \alpha-i\alpha \left(g_1 q_1+g_2 q_2\right)+\varepsilon-\frac{\kappa}{2}\alpha,\\
0&=&-m\omega_{\text{m}}^2 q_1-\gamma p_1-\hbar g_1 |\alpha|^2, \\
0&=&-m\omega_{\text{m}}^2 q_2-\gamma p_2-\hbar g_2 |\alpha|^2, \\
0&=&\frac{p_1}{m}, \quad 0=\frac{p_2}{m}.
\end{eqnarray}
Then, we have
\begin{eqnarray}
 q_1&=&\frac{\hbar g_1}{m\omega_m^2} |\alpha|^2, \quad q_2=\frac{\hbar g_2}{m\omega_m^2} |\alpha|^2, \quad p_1=p_2=0, \\
 \varepsilon&=& \alpha \left(i\Delta_0 +i \hbar\frac{g^2_1+g^2_2}{m\omega_m^2}|\alpha|^2 -\frac{\kappa}{2}\right),
\end{eqnarray}
which can be solve analytically to obtain $q_1$, $q_2$, and $\alpha$. Then, we have the dynamical matrix
\begin{equation}
A^{(TR)}=\begin{pmatrix}
 -\frac{\kappa}{2} & \Delta & 0 & 0 &\sqrt{2}g_1\text{Im}[\alpha] & \sqrt{2}g_2\text{Im}[\alpha] \\
 -\Delta & -\frac{\kappa}{2} & 0 & 0 & -\sqrt{2}g_1\text{Re}[\alpha] & -\sqrt{2}g_2\text{Re}[\alpha] \\
 -\sqrt{2}\hbar g_1\text{Re}[\alpha] & -\sqrt{2}\hbar g_1\text{Im}[\alpha] & -\gamma  & 0
   & -m \omega_m^2 & 0 \\
 -\sqrt{2}\hbar g_2\text{Re}[\alpha] &
  -\sqrt{2}\hbar g_2\text{Im}[\alpha] & 0 & -\gamma  & 0 & -m
   \omega_m^2 \\
 0 & 0 & \frac{1}{m} & 0 & 0 & 0 \\
 0 & 0 & 0 & \frac{1}{m} & 0 & 0
\end{pmatrix}
\end{equation}
acting on the vector of operators $(\hat{X},\hat{Y},\hat{p}_1,\hat{p}_2,\hat{q}_1,\hat{q}_2)^T$ with $\Delta=\Delta_0+\hbar\frac{g^2_1+g^2_2}{m\omega_m^2}|\alpha|^2$. In our numerical simulations 
the stability of $A^{(TR)}$ is always satisfied.

\section{Weyl transform of the SLD}
\label{app:WeylTransform}

In the main text, we have used the phase space formalism which relies on the Weyl transform \cite{Schleich}, a map from bounded operators to functions on the phase space. 
The Weyl transform $A(x,y)$ of an operator $\hat{A}$ is defined by
\begin{equation}\label{eq:WeylTransform}
A(x,y)=\int d\xi e^{-iy\xi}\langle x+\frac{\xi}{2}|\hat{A}|x-\frac{\xi}{2}\rangle.
\end{equation}

This approach is very useful for the calculation of the QFI for a Gaussian state \cite{SanavioBernadXuereb2020}, where the Weyl transform or Wigner function of a density operator $\hat{\rho}$ is
a Gaussian function. The SLD operator $\hat{\mathcal{L}}_i$ satisfies the relation \eqref{eq:SLDrelation} and for a Gaussian state its Weyl transform corresponds to the expression 
in eq.~\eqref{eq:WeylSLD}, or explicitly

\begin{eqnarray}
L^i(x,y) &=& \Phi^i_{11}x^2+\Phi^i_{22}y^2+2\Phi^i_{12}xy-\nu^i.
\end{eqnarray}

The inverse transformation of this function yields the following operator 
\begin{eqnarray}
\hat{\mathcal{L}}_i &=& \Phi^i_{11}\hat{x}^2+\Phi^i_{22}\hat{y}^2+\Phi^i_{12}(\hat{x}\hat{y}+\hat{y}\hat{x})-\nu^i\mathds{1},
\end{eqnarray}
where one has to use the Weyl-ordering.

Now, the Weyl transform ~\eqref{eq:WeylTransform} is applied on the operator $\hat{\mathcal{L}}_i\hat{\mathcal{L}}_j$ yielding

\begin{eqnarray}
L^{(2)}_{ij}(x,y)&=&\Phi_{11}^i\Phi_{11}^jx^4+2(\Phi_{11}^i\Phi_{12}^j+\Phi_{11}^j\Phi_{12}^i)x^3y+(\Phi_{11}^i\Phi_{22}^j+\Phi_{11}^j\Phi_{22}^i+4\Phi_{12}^i\Phi_{12}^j)x^2y^2 \nonumber\\
&+&2(\Phi_{22}^i\Phi_{12}^j+\Phi_{22}^j\Phi_{12}^i)xy^3+\Phi_{22}^i\Phi_{22}^jy^4-(\Phi_{11}^i\nu^j+\Phi_{11}^j\nu^i)x^2-(\Phi_{22}^i\nu^j+\Phi_{22}^j\nu^i)y^2\nonumber\\
&-&\frac{1}{2}(\Phi_{11}^i\Phi_{22}^j+\Phi_{22}^i\Phi_{11}^j-2\Phi_{12}^i\Phi_{12}^j)+\nu^i\nu^j.
\end{eqnarray}

The QFI matrix entries are the mean values of the above set of functions with $i,j \in \{1,2\}$, which are calculated by integrating them with the Gaussian Wigner function $W(x,p)$. 
This leads to eq. \eqref{eq:QFIcomponentsWT}. 

\end{widetext}

\end{document}